\documentclass[a4paper,fleqn,usenatbib]{mnras}

\usepackage{mathptmx}

\usepackage[T1]{fontenc}
\usepackage{ae,aecompl}

\usepackage{graphicx}	
\usepackage{amsmath}	
\usepackage{amssymb}	
\usepackage{multirow, array}

\newcommand{\fili}{\textit{i'}}
\newcommand{\fr}{\textit{r'}}
\newcommand{\fg}{\textit{g'}}
\newcommand{\fu}{\textit{u'}}
\newcommand{\fKG}{\textit{KG5}}

\newcommand{\warwick}{{1}}
\newcommand{\sheffield}{{8}}
\newcommand{\surrey}{{5}}
\newcommand{\dpmacal}{{2}}
\newcommand{\radboud}{{3}}
\newcommand{\ioacambs}{{4}}
\newcommand{\narit}{{6}}
\newcommand{\ljm}{{7}}
\newcommand{\iac}{{9}}
\newcommand{\cahill}{{10}}
\newcommand{\unc}{{11}}
\newcommand{\jpl}{{12}}

\newcommand{\review}[1]{{#1}}

\title[High-Speed Photometry of Gaia14aae]{High-Speed Photometry of Gaia14aae: An Eclipsing AM\,CVn That Challenges Formation Models}

\author[M. J. Green et al.]{M. J. Green,$^\warwick$\thanks{E-mail: matthew.green@warwick.ac.uk (MJG)}
T. R. Marsh,$^\warwick$
D. T. H. Steeghs,$^\warwick$
T. Kupfer,$^\dpmacal$
R. P. Ashley,$^\warwick$
\newauthor
S. Bloemen,$^\radboud$
E. Breedt,$^{\warwick,\ioacambs}$
H. C. Campbell,$^{\ioacambs,\surrey}$
A. Chakpor,$^\narit$
C. M. Copperwheat,$^\ljm$
\newauthor
V. S. Dhillon,$^{\sheffield, \iac}$ 
G. Hallinan,$^\cahill$
L. K. Hardy,$^\sheffield$
J. J. Hermes,$^\unc$\thanks{Hubble Fellow}
P. Kerry,$^\sheffield$
\newauthor
S. P. Littlefair,$^\sheffield$ 
J. Milburn,$^\cahill$
S. G. Parsons,$^\sheffield$
N. Prasert,$^\narit$
J. van Roestel,$^\radboud$
\newauthor
D. I. Sahman,$^\sheffield$
and N. Singh.$^{\cahill, \jpl}$
\\
$^\warwick$Astronomy and Astrophysics Group, Department of Physics, University of Warwick, Coventry, CV4 7AL, United Kingdom
\\
$^\dpmacal$Division of Physics, Mathematics and Astronomy, California Institute of Technology, Pasadena, CA 91125, USA
\\
$^\radboud$Department of Astrophysics/IMAPP, Radboud University, PO Box 9010, NL-6500 GL Nijmegen, the Netherlands
\\
$^\ioacambs$Institute of Astronomy, University of Cambridge, Madingley Road, Cambridge, CB3~0HA, United Kingdom
\\
$^\surrey$Department of Physics, University of Surrey, Guildford, GU2 7XH, United Kingdom
\\
$^\narit$National Astronomical Research Institute of Thailand (Public Organization), 260 Moo 4, T. Donkaew, A. Maerim, \\Chiangmai 50180, Thailand
\\
$^\ljm$Astrophysics Research Institute, Liverpool John Moores University, Liverpool L3 5RF, UK
\\
$^\sheffield$Department of Physics and Astronomy, University of Sheffield, Sheffield, S3~7RH, United Kingdom
\\
$^\iac$Instituto de Astrof\'isica de Canarias, 38205 La Laguna, Tenerife, Spain
\\
$^\cahill$Cahill Centre for Astronomy and Astrophysics, California Institute of Technology, Pasadena, CA 91125, USA
\\
$^\unc$Department of Physics and Astronomy, University of North Carolina, Chapel Hill, NC 27599-3255, USA
\\
$^\jpl$Jet Propulsion Laboratory, California Institute of Technology, Pasadena, CA 91109, USA
}

\date{Accepted XXX. Received YYY; in original form ZZZ}

\pubyear{2017}

\begin{document}
\label{firstpage}
\pagerange{\pageref{firstpage}--\pageref{lastpage}}
\maketitle

\begin{abstract}
AM\,CVn-type systems are ultra-compact, hydrogen-deficient accreting binaries with degenerate or semi-degenerate donors. The evolutionary history of these systems can be explored by constraining the properties of their donor stars. We present high-speed photometry of Gaia14aae, an AM\,CVn with a binary period of 49.7~minutes and the first AM\,CVn in which the central white dwarf is fully eclipsed by the donor star. 
Modelling of the lightcurves of this system allows for the most precise measurement to date of the donor mass of an AM\,CVn, and relies only on geometric and well-tested physical assumptions. We find a mass ratio $q = M_2/M_1 = 0.0287 \pm 0.0020$ and masses $M_1 = 0.87 \pm 0.02 M_\odot$ and $M_2 = 0.0250 \pm 0.0013 M_\odot$. We compare these properties to the three proposed channels for AM\,CVn formation. 
Our measured donor mass and radius do not fit with the contraction that is predicted for AM\,CVn donors descended from white dwarfs or helium stars at long orbital periods. 
The donor properties we measure fall in a region of parameter space in which systems evolved from hydrogen-dominated cataclysmic variables are expected, but such systems should show spectroscopic hydrogen, which is not seen in Gaia14aae. 
The evolutionary history of this system is therefore not clear. We consider a helium-burning star or an evolved cataclysmic variable to be the most likely progenitors, but both models require additional processes and/or fine-tuning to fit the data. 
Additionally, we calculate an updated ephemeris which corrects for an anomalous time measurement in the previously published ephemeris. 
\end{abstract}




\section{Introduction}
\label{Introduction}

AM\,CVn-type systems are a class of compact binaries in which white dwarfs accrete helium-dominated matter from low-mass degenerate or semi-degenerate companions. Their short orbital periods (5 -- 65 minutes) and deficiency of hydrogen make them probes of extreme physics, and they will be among the first systems detected by a space-based gravitational wave interferometer \citep{Korol2017,Nelemans2003}. Alongside the menagerie of hydrogen-dominated accreting white dwarfs (Cataclysmic Variables, CVs) they are laboratories of accretion physics \citep{Kotko2012,Cannizzo2015a}. They experience dwarf nova outbursts, and are evolutionarily related to the double-degenerate pathway towards type Ia supernovae. They are also proposed sources of subluminous `.Ia' supernovae \citep{Bildsten2007} and may be sources of helium novae \citep[such as V445 Pup,][]{Woudt2009}. 
AM\,CVns are rare in comparison to hydrogen-rich CVs, but the number of known systems has increased dramatically in recent years due to transient surveys \citep[eg.][]{Levitan2011,Levitan2013} and follow-up surveys based on colour selection from the Sloan Digital Sky Survey \citep[SDSS;][]{Roelofs2007,Carter2013a,Carter2014}. Around 50 systems are now known, twice the number quoted in the most recent review \citep{SolheimAMCVn}. However, the space density of the AM\,CVn population continues to fall short of predictions based on binary population synthesis models \citep{Carter2013a}.

AM\,CVns are the end-point of a finely tuned evolutionary process involving either one or two common envelope phases. As such, they provide an opportunity to calibrate models of interacting binary evolution, including the poorly-understood common-envelope phase \citep{Ivanova2013}.
There are three proposed channels by which AM\,CVn binaries may form, but the contribution of each channel to the AM\,CVn population is poorly constrained.

In the white dwarf channel \citep{Paczynski1967} and the helium-star channel  \citep{Savonije1986, Iben1987}, the binary passes through two common-envelope stages as each of its component stars leaves the main sequence. During these stages the binary is surrounded by an envelope of material that extracts energy from the system, reducing the orbital period of the system to below the period minimum of hydrogen-dominated, non-magnetic cataclysmic variables (CVs). The channels differ in the nature of the secondary star following the ejection of the second common envelope; in the white dwarf channel it is left as a low-mass, degenerate or semi-degenerate white dwarf, whereas in the helium-star channel it is a non-degenerate helium-burning star.
\citet{Deloye2007} and \citet{Yungelson2008} predict that donors from both the white dwarf and helium-star channels should evolve towards complete degeneracy at periods $\gtrsim$~40~minutes. For periods longer than this, the predicted donor mass for both channels is almost a unique function of orbital period alone. A similar convergence is predicted for helium white dwarf donors in ultra-compact X-ray binaries, a class of object with similar evolutionary paths \citep{Sengar2017}. 

These two channels are predicted by some models to dominate the formation of AM\,CVns at short periods \citep[<~25~minutes,][]{Nelemans2004}. On the other hand, \citet{Shen2015a} argues that friction within the ejecta of He novae could cause all double white dwarf binaries to merge directly rather than reaching a state of stable accretion, rendering the white dwarf donor channel essentially impossible. 

The third formation channel is the evolved-CV, or evolved main sequence donor, channel \citep{Tutukov1985, Podsiadlowski2003,Goliasch2015}. In this channel, the donor must evolve off the main sequence at around the time of the start of mass transfer. The system then appears as a hydrogen-dominated CV in its early evolution, becoming helium-dominated as the hydrogen envelope of the donor is stripped. This channel favours the formation of longer-period AM\,CVns (>~40~minutes). 
In addition to AM\,CVns, this channel is predicted to produce evolved CVs which contain both hydrogen and helium and have periods below the CV period minimum. 
Several CVs in the 50-76~minute period range have indeed been found which are possibly products of this channel \citep[eg.][]{Augusteijn1996,Breedt2012}. 
However, the small number of such systems compared to the number of AM\,CVns suggests that the channel should not be a major contributor to the AM\,CVn population.
This is consistent with population synthesis predictions which predict the formation of only a small number of AM\,CVns via this channel, due to the finely-tuned starting parameters and long timescales required to remove all visible hydrogen from these systems \citep[eg.][]{Goliasch2015}. 

Population estimates for these formation channels are poorly constrained by data due to the difficulty in distinguishing the products of these channels from one another. Products of the white dwarf and helium-star channels can be distinguished by the nature of their donor stars, as these channels produce donors with different levels of degeneracy which hence occupy different regions of a mass-radius diagram \citep{Deloye2007,Yungelson2008}. Due to the faintness of the donor star it cannot be observed directly. For many systems it can only be studied by indirect methods such as measuring the radial velocity of the accretor \citep[eg.][]{Kupfer2016a,Roelofs2006}, which generally does not yield results with the required precision to distinguish between the two channels, or methods based on the superhump period \citep[eg.][]{Kato2013} which arise from models that have not been well tested for helium-dominated systems.

In hydrogen CVs, high-speed photometry of eclipsing systems has yielded the most precise measurements of component masses and radii while relying only on assumptions about the geometry of the system.  \citep[eg. ][]{Savoury2011,Littlefair2014,McAllister2015}. Non-magnetic CVs have a geometry consisting of several key components: the donor and accretor stars, an accretion disc around the accretor, a stream of matter passing from the donor to the accretor, and a `bright spot' (sometimes referred to as a `hot spot') on the edge of the accretion disc at the point of intersection with the infalling matter stream. During eclipse these components add characteristic features to the lightcurve which can be used to constrain the properties of the system \citep{cook1984,Wood1986a}.

Only three eclipsing AM\,CVns have been discovered to date: PTF1J1919+4815, in which just the edge of the disc and bright spot are eclipsed \citep{Levitan2014}; YZ\,LMi (also known as SDSS~J092638.71+362402.4), in which the white dwarf is partially eclipsed \citep{Anderson2005,Copperwheat2011}; and Gaia14aae (also known as ASASSN-14cn), the only known AM\,CVn in which the white dwarf is fully eclipsed \citep{Campbell2015}. Due to their eclipsing nature, the latter two of these systems are ideal targets for parameter studies. \citet{Copperwheat2011} used eclipse fitting with high time-resolution photometry to measure the donor mass of YZ\,LMi. This paper does the same for Gaia14aae.

\begin{table*}
	\centering
	\caption{Summary of the observations carried out for this paper.}
	\label{tab:observations}
	\begin{tabular}{lcccr} 
		\hline
		Observatory & Date at start (UT) & Filters & Exposure time (s) * & \begin{tabular}{@{}r@{}}Num.\\Eclipses\end{tabular}\\
		\hline
		WHT + ULTRACAM & 2015 01 14 & \textit{u'g'i'} & 3 (9) & 3\\
		& 2015 01 15 & \textit{u'g'r'} & 3 (9) & 4\\
		& 2015 01 16 & \textit{u'g'r'} & 3 (9) & 5\\
		& 2015 01 17 & \textit{u'g'r'} & 3 (9) & 3\\
		& 2015 05 23 & \textit{u'g'r'} & 2.5 (7.5) & 6\\
		& 2015 06 22 & \textit{u'g'r'} & 3 (9) & 4\\
		TNT + ULTRASPEC & 2016 03 12 & \textit{KG5} & 8 & 1\\
		 & 2016 03 13 & \textit{KG5} & 5 & 2\\
		 & 2016 03 14 & \textit{KG5} & 5 & 3\\
		 & 2016 03 15 & \textit{g'} & 5 & 2\\
		Hale + CHIMERA & 2016 08 06 & \textit{g'r'} & 4 & 7\\
		 & 2016 08 07 & \textit{g'r'} & 4 & 6\\
		 & 2016 08 08 & \textit{g'r'} & 4 & 6\\
		TNT + ULTRASPEC & 2017 02 21 & \textit{KG5} & 4 & 1\\
		\hline
	\multicolumn{5}{p{10cm}}{* Brackets denote the {\fu}  exposure time, 
	which was increased by a factor of 3 to compensate for the 
	lower sensitivity in that band.}\\
	\hline
	\end{tabular}
\end{table*}

Gaia14aae has a {\fg} magnitude of 18.3--18.6 outside of eclipse and an orbital period of 49.71 minutes, putting it at the long-period end of the AM\,CVn distribution. 
Gaia14aae experienced two outbursts in June and August 2014, but no outbursts have been recorded since then. Following the discovery of Gaia14aae, \citet{Campbell2015} used time-series photometry to constrain the properties of the system. They were not able to measure the mass ratio, but were able to set a lower limit of $q > 0.019$, and found corresponding minimum masses for the primary and secondary of 0.78 and 0.015 $M_{\odot}$ respectively. 



This paper presents follow-up photometry and analysis of Gaia14aae. In Section~\ref{Observations} we will describe how the data were taken and reduced. Section~\ref{Data} will present the data. Section~\ref{Analysis} will describe the process used to fit models to the eclipses and present the results. In Section~\ref{Discussion} we will compare these results to modelled evolutionary tracks.

\section{Observations}
\label{Observations}

In this paper we present high-speed, multi-colour photometric observations of Gaia14aae, taken using the high-speed CCD cameras ULTRACAM, ULTRASPEC and CHIMERA. We observed 53 eclipses, each in 1--3 colour bands, spanning a time period of 25 months. A summary of observations is presented in Table~\ref{tab:observations}.

Images were taken using ULTRACAM \citep{ultracam} on the 4.2~m William Herschel Telescope (WHT). ULTRACAM is a three-beam camera, allowing it to record images in three colour bands simultaneously, while using frame-transfer CCDs to reduce dead-time between exposures to negligible amounts (25~ms). Data were also collected using ULTRASPEC \citep{ultraspec}, a single-band photometer with a frame-transfer, electron-multiplying CCD and only 15~ms dead time per cycle, which is mounted on the 2.4~m Thai National Telescope (TNT). Most of the ULTRASPEC data were taken using a custom filter, {\fKG}, which has a broader band to allow for shorter exposure times given the smaller collecting area of this telescope \citep[for more details on the {\fKG} band, see][]{Hardy2017}. A further series of eclipses were observed using CHIMERA \citep{chimera}, a 2-band photometer which uses frame-transfer, electron-multiplying CCDs to achieve 15~ms dead time, and is mounted on the Hale 200-inch (5.1~m) Telescope.

The images from all three instruments were reduced using the ULTRACAM reduction pipeline, described in \citet{ultracam}. Each of the images were bias-subtracted and divided by twilight flat fields. The ULTRACAM and ULTRASPEC data were also dark-subtracted. The flux of Gaia14aae was then extracted using aperture photometry, dividing the flux in each frame by a comparison star (J2000 coordinates 16:11:30.53, +63:09:25.8) to remove any atmospheric transparency variations. 

The \textit{u'g'r'i'} fluxes were then calibrated using the comparison star, the magnitude of which is available from SDSS ($m_\fu$ = 17.52, $m_\fg$ = 15.45, $m_\fr$ = 14.62, $m_\fili$ = 14.34). We ensured that this star did not show variability and we tested this calibration using several nearby comparison stars, each of which resulted in consistent calibrations. Although the {\fKG} band can be flux-calibrated \citep[see the Appendix of][]{Hardy2017}, we did not calibrate these data because they are used for timing purposes only.

\section{Photometry}
\label{Data}

\begin{figure*}
	\includegraphics[width=500pt]{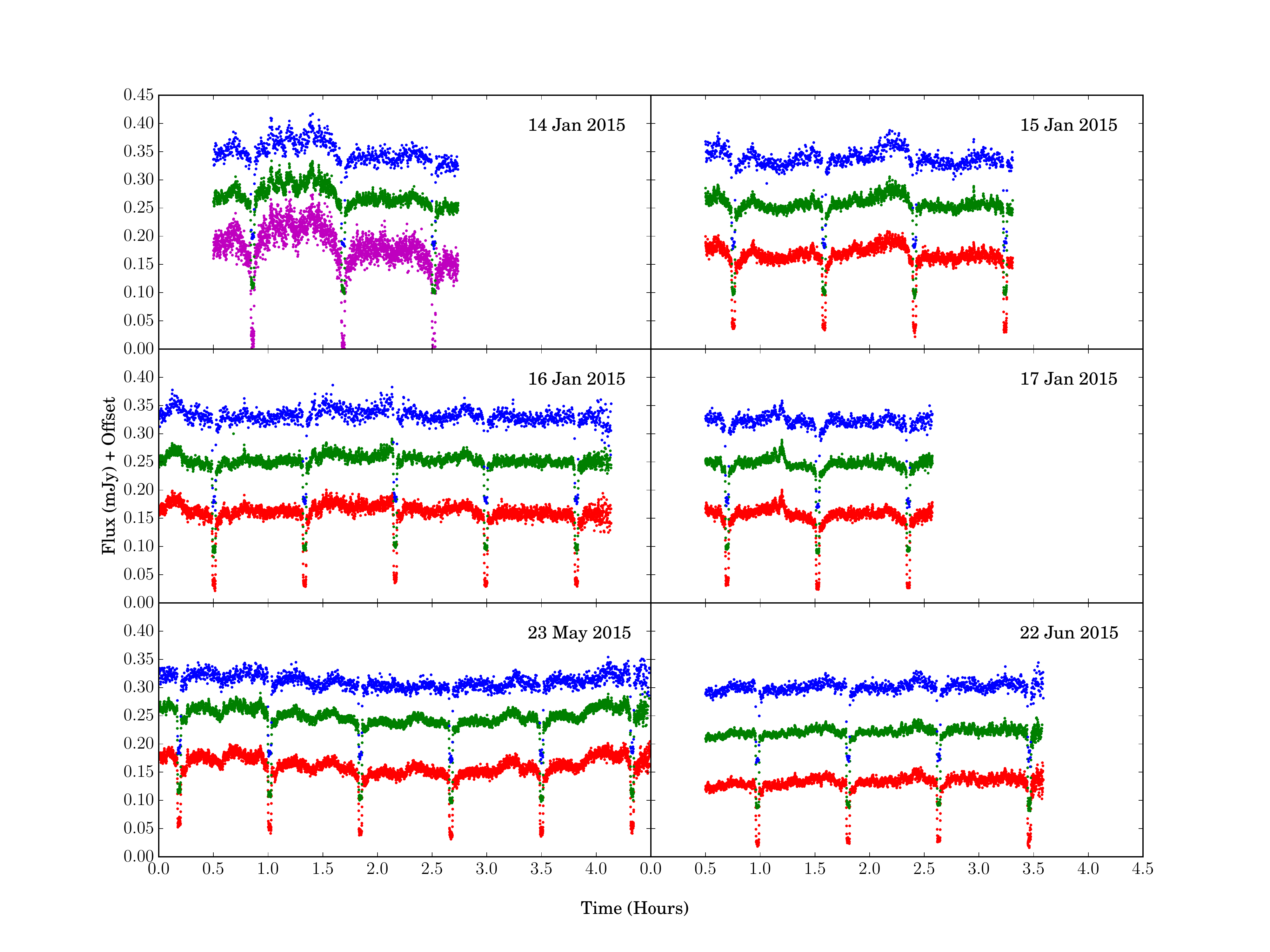}
    \caption{ULTRACAM photometry taken on the 4.2m WHT in 2015. Data shown were taken in {\fu} (blue), {\fg} (green), {\fr} (red), and {\fili} (magenta). The {\fili}, {\fg}, and {\fu} data have been offset in the y-direction by -0.07~mJy, 0.07~mJy, and 0.15~mJy, respectively. No offset has been applied to the {\fr} data. Data with error bars more than 3.5 times the mean (due to cloud) have been removed for clarity.
    }
    \label{fig:all-photom1}
\end{figure*}

\begin{figure*}
	\includegraphics[width=500pt]{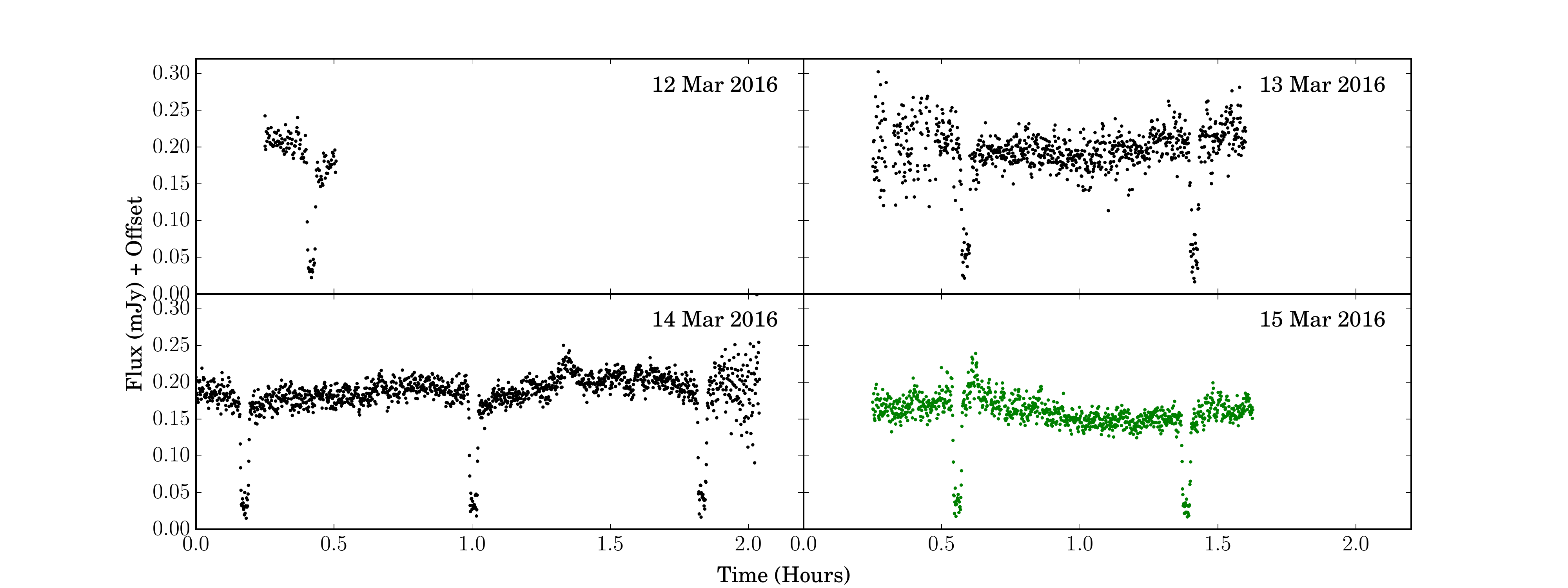}
    \caption{ULTRASPEC photometry taken on the 2.4m TNT in 2016. Data were taken in {\fKG} (black) and {\fg} (green). Note that the scaling of the {\fKG} data is arbitrary as those data are not flux-calibrated. Data with error bars more than 3.5 times the mean (due to cloud) have been removed for clarity.
    }
    \label{fig:all-photom2}
\end{figure*}

\begin{figure*}
	\includegraphics[width=500pt]{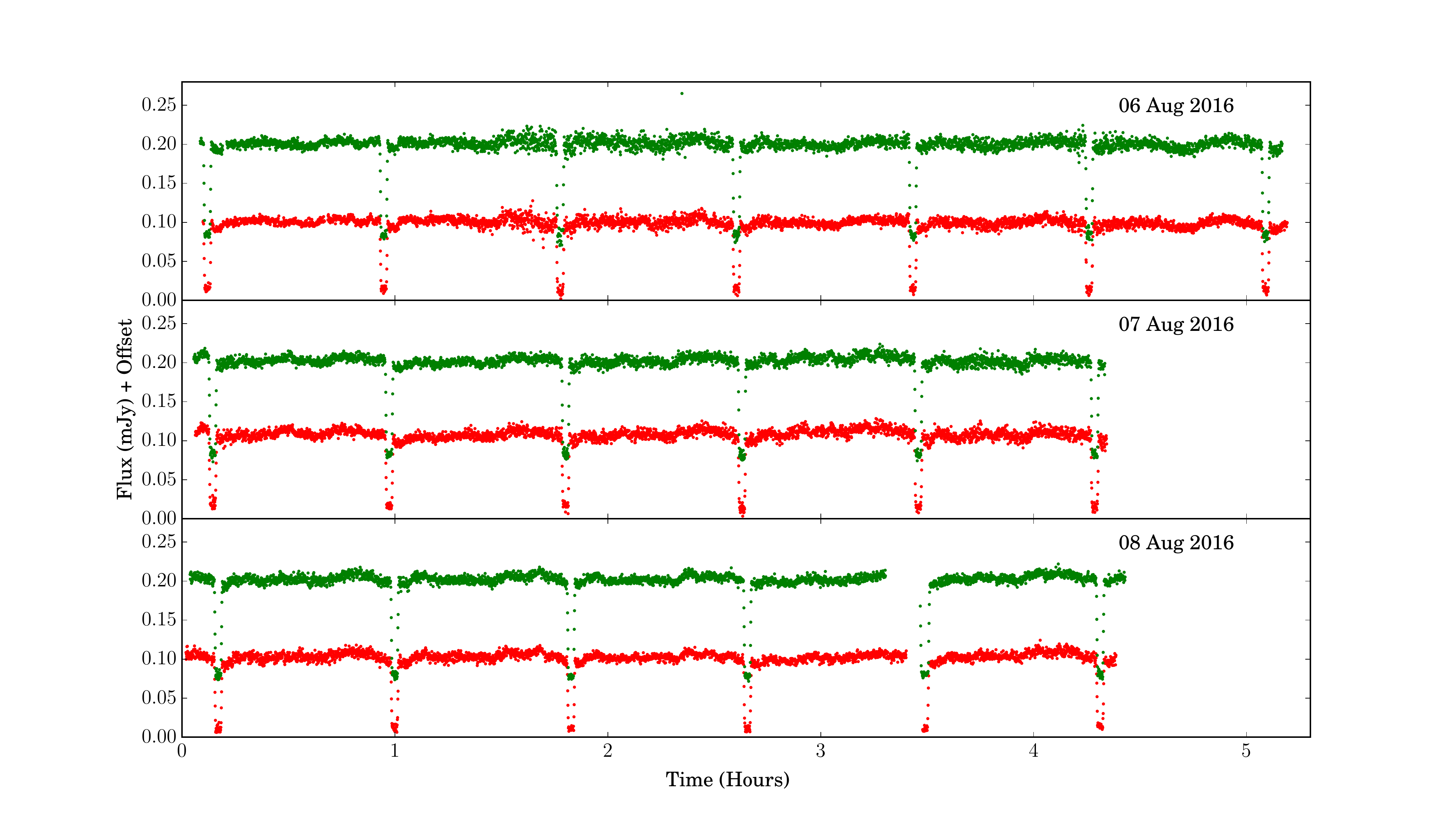}
    \caption{The CHIMERA data taken on the 5.1m Hale telescope in August 2016. These data were taken in {\fg} (green) and {\fr} (red). The {\fg} data have been offset by 0.07~mJy. Data with error bars more than 3.5 times the mean (due to cloud) have been removed for clarity.
    }
    \label{fig:all-photom3}
\end{figure*}

\begin{figure*}
	\includegraphics[width=500pt]{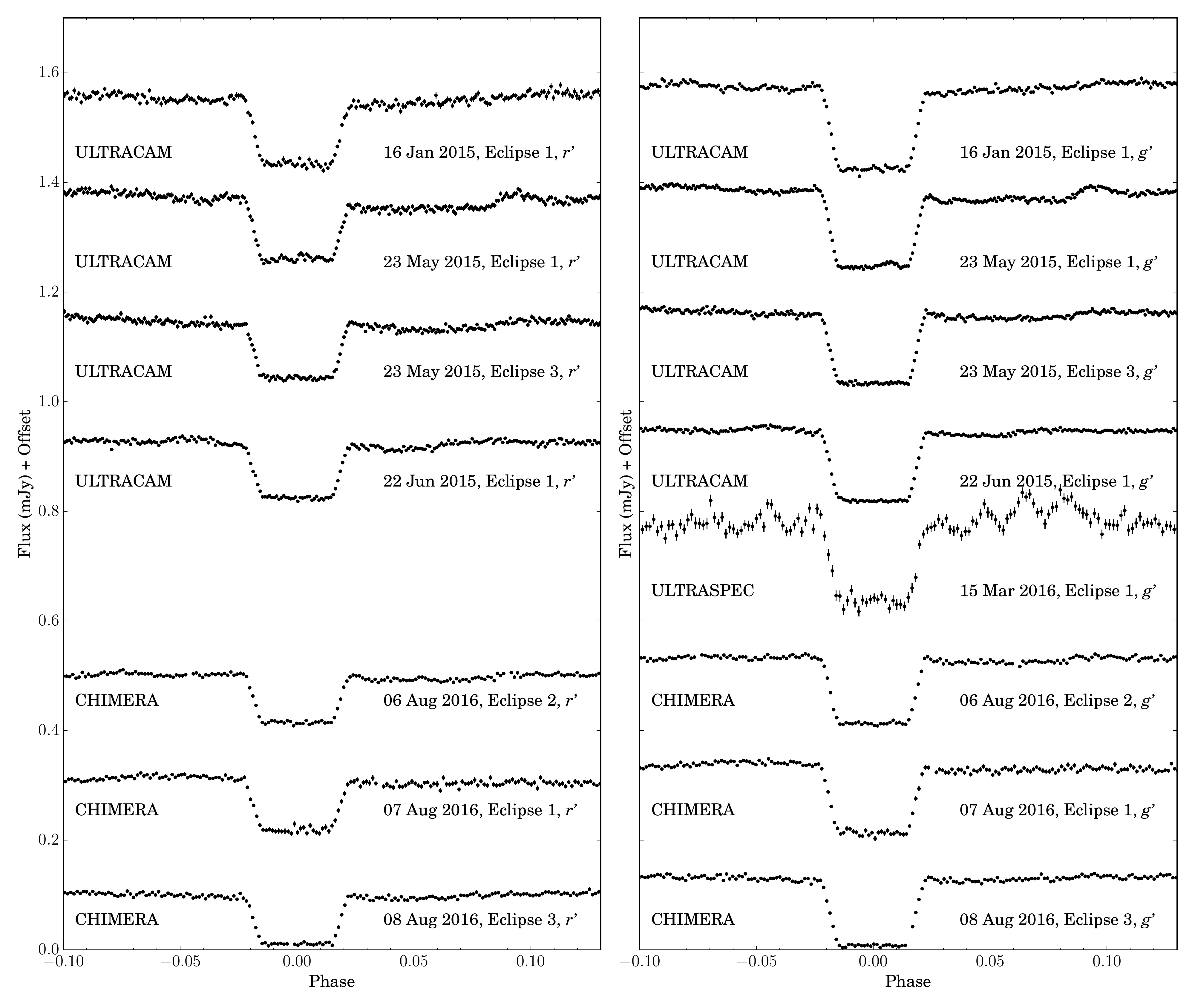}
    \caption{Example lightcurves of Gaia14aae, demonstrating the variability of the eclipse shape. Each lightcurve has been offset by 0.2~mJy from the lightcurve below it.
    }
    \label{fig:selected-eclipses}
\end{figure*}

\begin{figure*}
	\includegraphics[width=\columnwidth]{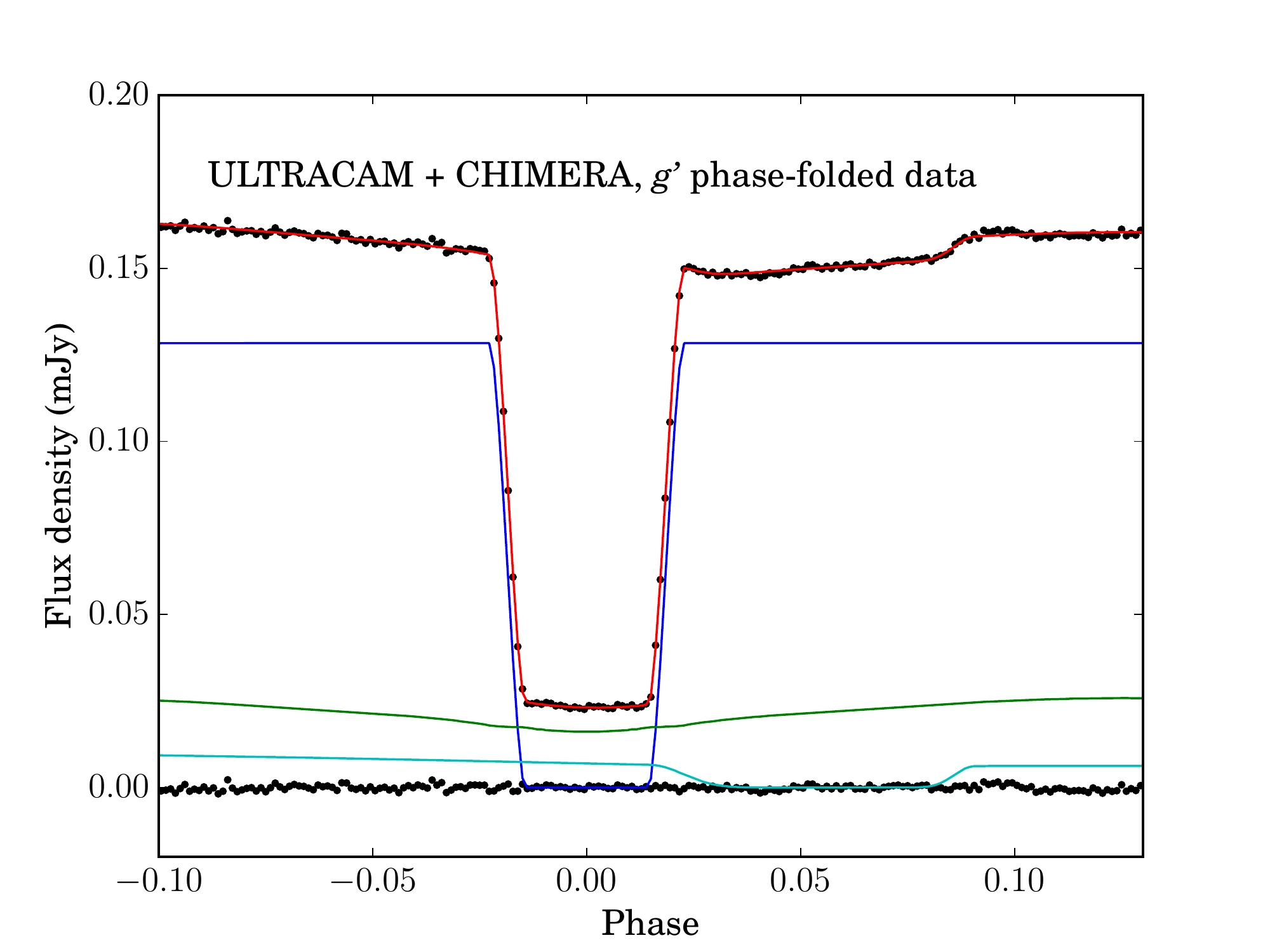}
	\includegraphics[width=\columnwidth]{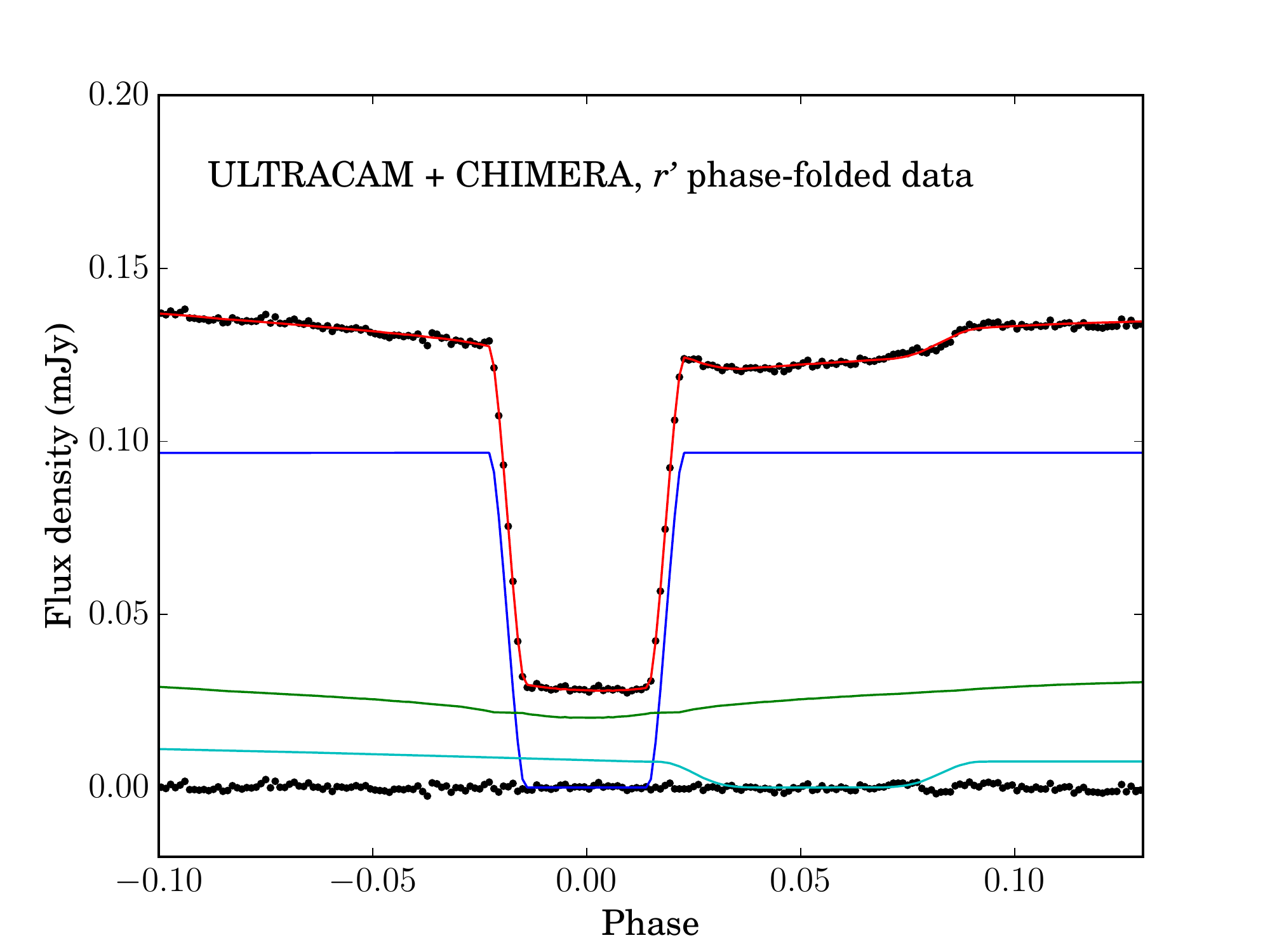}
    \caption{Phase-folded, rescaled, and averaged light curves combining both ULTRACAM and CHIMERA data, in {\fg} and {\fr} (left and right respectively, black points). The lightcurves show the consecutive eclipses of the accreting white dwarf and the bright spot. We show our best-fit models to each (red line) and the residuals. We also show a breakdown of the lightcurve contributions of each component: central white dwarf (blue), accretion disc (green), and bright spot (cyan). The donor's contribution was assumed to be negligible. There is a feature following the egress of the bright spot (phase 0.1) which is not described by the model, and may be due to flickering.
    }
    \label{fig:averaged-eclipses}
\end{figure*}

The complete set of photometry is shown in Figures~\ref{fig:all-photom1}-\ref{fig:all-photom3}, and Figure~\ref{fig:selected-eclipses} shows an example set of individual eclipses. 
Figure~\ref{fig:averaged-eclipses} shows \textit{r'}- and \textit{g'}-band phase-folded eclipses.

The flux from Gaia14aae is dominated by the central white dwarf, with the accretion disc and bright spot also making measurable contributions. Due to its low temperature, the contribution of the donor is not seen. 
Figure~\ref{fig:averaged-eclipses} shows a decomposition of the contributions of each component to the eclipse profile. The bright spot enters eclipse just as the white dwarf emerges, weakening the white dwarf's egress feature and causing an asymmetry in its eclipse shape.
Some lightcurves (Figure~\ref{fig:all-photom1}) also show a periodic `hump' feature around phases -0.5~to~0, which is commonly seen in CVs and which indicates that the light from the bright spot is preferentially emitted in a certain direction rather than being isotropic.

Gaia14aae shows the short-term variability known as `flickering' which is commonly observed in accreting systems. This arises from the variable nature of the accretion, and is generally localised to the bright spot and the inner disc. Flickering provides a source of correlated noise that is difficult to deal with analytically, particularly as the amplitude of the variability changes throughout the cycle with the eclipse of the bright spot and inner disc \citep{McAllister2017}. In Gaia14aae, the amplitude of the flickering is similar to the depth of the bright spot eclipse, meaning that in some cycles the bright spot eclipse is completely hidden by the flickering (see some of the examples in Figure~\ref{fig:selected-eclipses}). The amplitude of flickering in Gaia14aae varies significantly between epochs, with the system being particularly variable in some epochs (eg. March 2016, Figure~\ref{fig:all-photom2}) and comparatively stable in others (eg. August 2016, Figure~\ref{fig:all-photom3}).

Alongside this flickering, Gaia14aae shows a significant amount of variability on longer timescales. The total flux from the system decreases by approximately $30\%$ between the earliest and latest observations, with no corresponding colour change. Nights such as 23~May~2015 (Figure~\ref{fig:all-photom1}) show a long-term variation in flux. 
The depths of both the bright spot eclipse and the disc eclipse change significantly throughout the period of observations, both being particularly prominent in January 2015 (Figure~\ref{fig:all-photom1}) and barely visible by August 2016 (Figure~\ref{fig:all-photom3}). The phase of the bright spot ingress also appears later in the 2016 data than in the 2015 data, which may imply an increase in the accretion disc radius. This is somewhat unexpected as the disc radius is usually expected to decrease with time during periods of quiescence \citep[eg.][]{Wood1989a}. The depth of the white dwarf eclipse shows a significant decrease over the time period of observation (see Section~\ref{luminosity} for further analysis of the white dwarf eclipse depth). 

\section{Light Curve Modelling}
\label{Analysis}

The lightcurves contain the necessary information to derive the donor mass of the system \citep{cook1984,Wood1986a}, given an assumed $M$--$R$ relation for the central white dwarf. If the donor is assumed to be Roche-lobe filling, its radius as a fraction of orbital separation, $R_2/a$, is a function of the mass ratio, $q$, of the two stars. 
The phase width of the white dwarf eclipse $\Delta \phi$ therefore depends only on $q$ and the orbital inclination $i$.
For a given $\Delta \phi$, the radius of the donor (and hence $q$) can be increased by moving the system towards lower orbital inclinations (more face-on). A lower limit of $q$ can be found from $\Delta \phi$ by assuming $i=90^{\circ}$. Using this method, \citet{Campbell2015} found $q>0.019$ for Gaia14aae.

Lifting the degeneracy between $q$ and $i$ requires additional information, which can be found from the contact phases of the bright spot eclipse. The path of the infalling stream of matter relative to the two stars (on which the bright spot is assumed to lie) is a function of $q$. The bright spot ingress and egress phases therefore provide an additional constraint on $q$ and $i$ that serves to disentangle them. 

Once $q$ and $i$ are known, the radius $R_1$ of the accreting white dwarf can be found from the duration of the ingress or egress of the white dwarf eclipse $\Delta w$. An assumed $M$--$R$ relation for the accretor can then give $M_1$, and hence $M_2$ can be found from $q$.

\subsection{MCMC Modelling}

We modelled the eclipses of Gaia14aae using the package \texttt{lcurve}. A description of the model can be found in \citet[Appendix]{Copperwheat2010}, but we give a brief overview here. The code models each component of the system (white dwarf, accretion disc, bright spot and donor star) as a grid of elements, and for each timestep computes which elements are occulted by the donor star. The white dwarf is assumed to be spherical, the disc to be 2D and azimuthally symmetric, and the donor to be Roche-lobe filling. The bright spot is modelled as a 1D line of points, of whose light some fraction is emitted isotropically while the remainder is `beamed' at an angle from the system, allowing for the recreation of the hump feature in lightcurves. Key variables in the model are the mass ratio $q$, orbital inclination $i$, mid-time of the white dwarf eclipse $t_0$, orbital period $P_\text{orb}$, the relative temperatures of all components, the radius of the accretor, the outer radius of the disc, and the orientation, beamed fraction and angle of beaming of the bright spot. All variables are listed in Table~\ref{tab:lcurve}.
Limb-darkening of the accretor was described according to \citet{Gianninas2013c} for a 13000K, log($g$)=8.5 white dwarf.

\begin{table*}
	\centering
	\caption{Details on variables in \texttt{lcurve} and any constraints applied during the fit to the phase-folded lightcurves.}
	\label{tab:lcurve}
	\begin{tabular}{l c} 
		\hline
		Variable & Description\\
		\hline
		$t_0$ & Mid-time of the white dwarf eclipse \\
		$q$ & Mass ratio $M_2 / M_1$ \\
		$\Delta i$ & Departure of inclination $i$ from that expected given $q$ and the average eclipse width $\Delta \phi = 0.037$ \\
		$R_1$ & Radius of the white dwarf \\
		$R_\text{bright spot}$ & Radial distance of the spot from the centre of the white dwarf\\
		$R_\text{disc}$ & Radial distance of the disc edge from the centre of the white dwarf\\
		$T_\text{white dwarf}$ & \multirow{3}{40em}{\centering Effective temperatures of the components -- these are really scaling factors for the flux contributions and do not reflect the true temperatures of these components} \\
		$T_\text{bright spot}$ & \\
		$T_\text{disc}$ & \\
		Spot length & Scale factor for the line of elements that make up the bright spot \\
		Spot angle & Angle between the line of elements of the bright spot and a tangent to the edge of the circle \\
		\multirow{2}{6em}{Spot fraction} & \multirow{2}{35em}{\centering Fraction of light from the bright spot which is `beamed' in a certain direction, producing the observed hump in the lightcurves} \\
		&\\
		Spot yaw & Beaming angle of the light from the bright spot in the orbital plane \\
		\hline
	\end{tabular}
\end{table*}

In order to find the optimal values and uncertainties of these parameters, this model was converged on the data using MCMC code implemented by the Python package \texttt{emcee} \citep{Foreman-Mackey2013}. 
\review{This package implements an affine-invariant ensemble sampling algorithm \citep{Goodman2010} in which the parameter space is explored by a cloud of `walkers'. At each iteration, new positions for each walker are proposed according to the positions of other walkers in the ensemble. In this way, the number of tuning parameters is vastly reduced compared to Metropolis-Hastings algorithms. A `scale factor' is required to tune the scale of each move; we left this at its default value of 2.
}
For each parameter in each fit, the best-fit value was taken to be the median value from all chains after the burn-in phase, with 1$\sigma$ error bars taken as the standard deviation. 

In order to measure changes in white dwarf flux and disc flux between eclipses, each eclipse in each colour band was first fitted separately using a shorter MCMC run. The ULTRACAM data were converged on using 50 walkers and a minimum of 20000 trials. The CHIMERA data were converged on using 200 walkers and a minimum of 5000 trials. This was sufficient to measure the relative flux of each component. Values of $q$ or $i$ based on individual eclipses have large error bars, and some eclipses are biased by flickering, by the coincidence of white dwarf egress with bright spot ingress, or show too faint a bright spot eclipse for $q$ to be well constrained. However, they are sufficient to provide a warning if the fit to the phase-folded lightcurve were a long way from the true value, and to compare properties of the white dwarf and disc which are independent of the bright spot parameters.

To measure $q$ and $i$, we produced phase-folded, binned and averaged lightcurves using all ULTRACAM and CHIMERA data in the \textit{g'} and \textit{r'} bands. The ULTRASPEC data were excluded because these data have larger error bars, they were taken during poorer observing conditions, and the system showed more than the usual amount of flickering during that run. Data in {\fili} and {\fu} were excluded due to the smaller number of eclipses available for both, as well as the poorer cadence and noisier data in {\fu}.

To deal with the variability in both white dwarf eclipse depth and baseline brightness, the individual eclipses were scaled based on their brightness both in-eclipse (phase 0) and after-eclipse (phase 0.1 -- 0.5) so as to reduce the variability within each bin. In the model fitting described above, $q$ and $i$ are constrained primarily by the ingress and egress phases of the white dwarf and bright spot, and therefore should not be affected by changes in the flux from the system. 
The phases of the bright spot can change if the disc radius changes; however, as only a small change in the disc radius is seen, this is likely not to be significant compared to other sources of uncertainty (particularly the scatter in white dwarf radius discussed in Section~\ref{luminosity}).
These lightcurves were converged on with 200 walkers and a minimum of 10 000 trials.

From each fitted model we found a value for $M_2$ based on the measured values of $q$ and $R_1/a$, Kepler's laws, and a white dwarf $M$--$R$ relation for the central white dwarf \citep[that of P.~Eggleton as quoted in][]{Verbunt1988}. The $M$--$R$ relation was scaled by a factor of 0.9985, chosen as the ratio between the Eggleton radius and the stellar model radius for a DB white dwarf of temperature and mass similar to ours \citep{Holberg2006,Bergeron2011}. This scaling factor represents both the nonzero temperature of the white dwarf and the lack of hydrogen in its atmosphere. We assumed a white dwarf temperature of 12900K, matching that measured by \citet{Campbell2015}.

\subsection{Bootstrapping}

The flickering of the system and the variability of the white dwarf discussed may induce systematic errors in the lightcurve fits. This error is not taken into account in our fit to the phase-folded data. In order to estimate the effect of these possible systematic errors on the mass ratio, we performed a bootstrap analysis. This involved selecting with replacement 44 from the 44 ULTRACAM and CHIMERA \textit{g'} eclipses and 41 from the 41 \textit{r'} eclipses, then phase-folding and binning the lightcurves. The procedure was repeated 1000 times in each band. Each output lightcurve was converged on with an \texttt{lcurve} model, using a combination of simplex and Levenberg-Marquardt algorithms. The results had a mean of $q = 0.0281 \pm 0.0007$, where the uncertainty is the standard deviation of individual results.

\subsection{Contact Phase Measurements}
\label{Contact Phases}

\begin{figure}
	\includegraphics[width=\columnwidth]{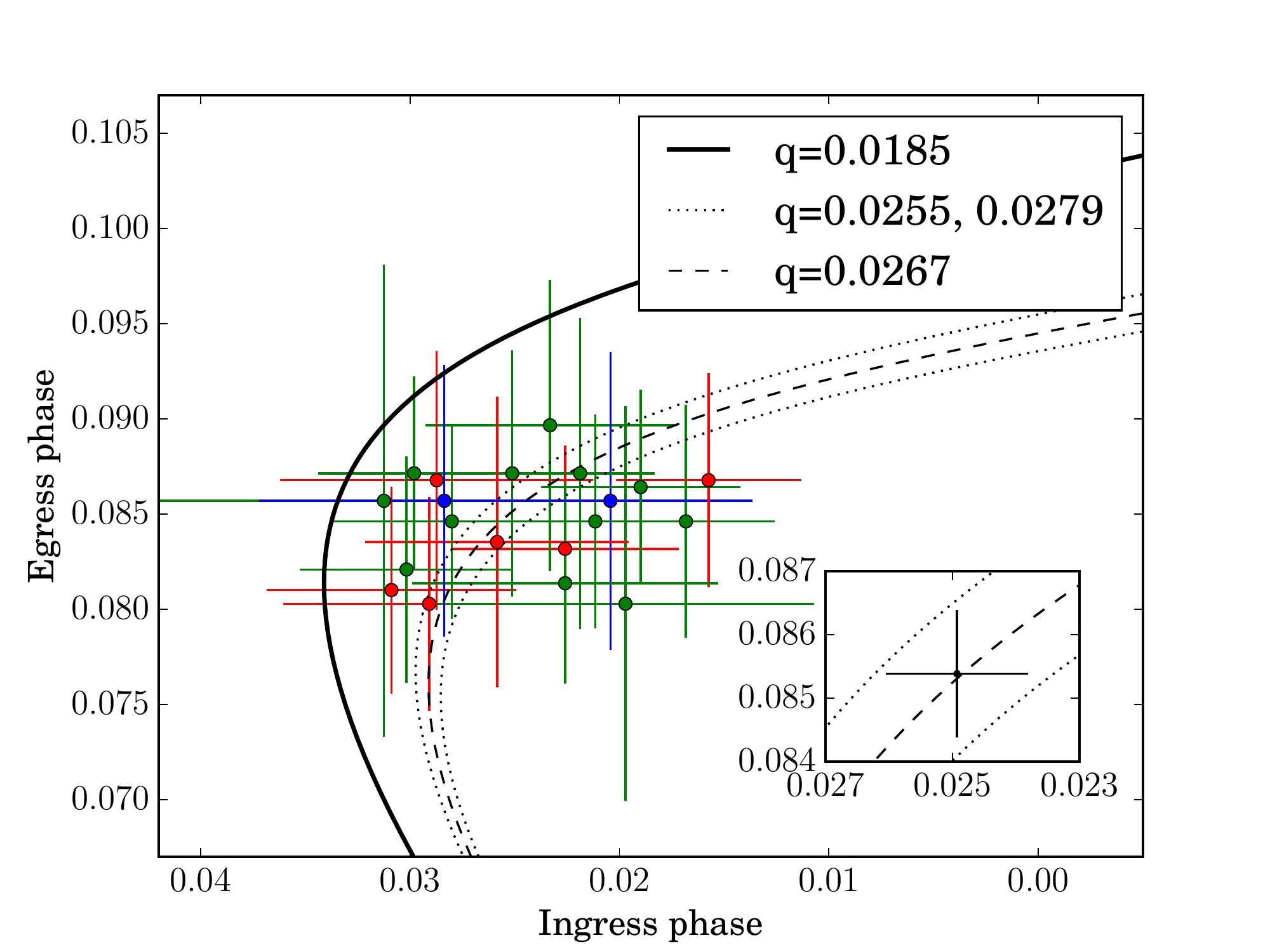}
    \caption{Bright spot ingress and egress phases measured by eye from the differential of individual ULTRACAM eclipse lightcurves\review{, using the method described in Section~\ref{Contact Phases}}. These measurements were made for {\fu} (blue), {\fg} (green) and {\fr} (red) eclipses. Also shown are ballistic stream paths in ingress/egress space for the $q$ values stated. In this manner $q$ can be estimated, under the assumption that the bright spot is a point source that lies on the ballistic stream path. The value $q=0.0185$ is the minimum mass ratio possible given the phase width of the white dwarf eclipse. \textit{Inset:} The weighted mean of these measurements, compared to the same ballistic streams.
    }
    \label{fig:ingress-egress}
\end{figure}

Several peculiarities of Gaia14aae make it particularly difficult to model by the method discussed above. Firstly, the bright spot of the system is weak when compared to the flickering of the system. Therefore, an inopportune spike of flickering during the bright spot eclipse can cause the fitting routine to misfit the bright spot feature, sending the model towards inaccurate parameter values. Secondly, the ingress of the bright spot eclipse coincides with the egress of the white dwarf eclipse, causing some amount of degeneracy in fitting these features.

We therefore explored an independent method of identifying $q$, in order to test the veracity of the MCMC modelling results. We carried out a method of measuring bright spot contact phases based on that described in \citet{Wood1986a}. From these contact phases a unique value of $q$ can be determined. Several assumptions, similar to those underlying the MCMC fitting method, are required for this: we assume that the bright spot is a point source, that it lies on the path of the infalling matter, and that the path of the matter can be described purely ballistically.

For each eclipse, we generated a model of best fit to the lightcurve. From this model we separated the white dwarf component and subtracted it from the original data, leaving residuals that should describe just the eclipses of the bright spot and accretion disc. We then performed a numerical differentiation on these residuals. In eclipses with clear bright spot features, the bright spot ingress is clearly visible in the differentiated data as a contiguous set of negative values, and the egress as positive values. Contact points 1, 2, 3, and 4 were defined as the start and end of ingress and the start and end of egress, respectively; in other words, the point at which the gradient departed from 0 or returned to 0. Mid-ingress was defined as halfway between contact points 1 and 2, and mid-egress as halfway between contact points 3 and 4. Uncertainties on each ingress and egress were judged by eye. Any eclipses without clear ingress or egress features were skipped.

Several caveats must be stated regarding this method \citep[see eg.][]{Feline2004}. Firstly, it is vulnerable to the same systematic biases as the MCMC method regarding flickering and the coincident white dwarf egress. Our hope is that the human eye might be better at identifying these features correctly than the automated technique, but this is far from certain. Secondly, it is subject to a human bias in that the observer looks for eclipse features where they are expected to be. We therefore use this method primarily as a check on the reasonableness of the MCMC results, rather than as an alternative method.

With the aforementioned caveats in place, the measured bright spot mid-ingress and mid-egress phases are shown in Figure~\ref{fig:ingress-egress}. There is a significant scatter in the results, which seems to be well-described by the estimated uncertainties. All measurements imply $q > 0.0185$, the minimum $q$ predicted from the phase width of the white dwarf eclipse (see Section~\ref{Discussion}). 
\review{The $1/\sigma$ weighted mean of these measurements is plotted in Figure~\ref{fig:ingress-egress}~(inset), along with uncertainties propagated from the error bars on individual measurements. These mean ingress and egress values and their uncertainties correspond to a value of $q = 0.0267 \pm 0.0012$.} This will be compared to the MCMC results given in Section~\ref{masses}.

\section{Modelling Results and Discussion}
\label{Discussion}

\begin{figure}
	\includegraphics[width=\columnwidth]{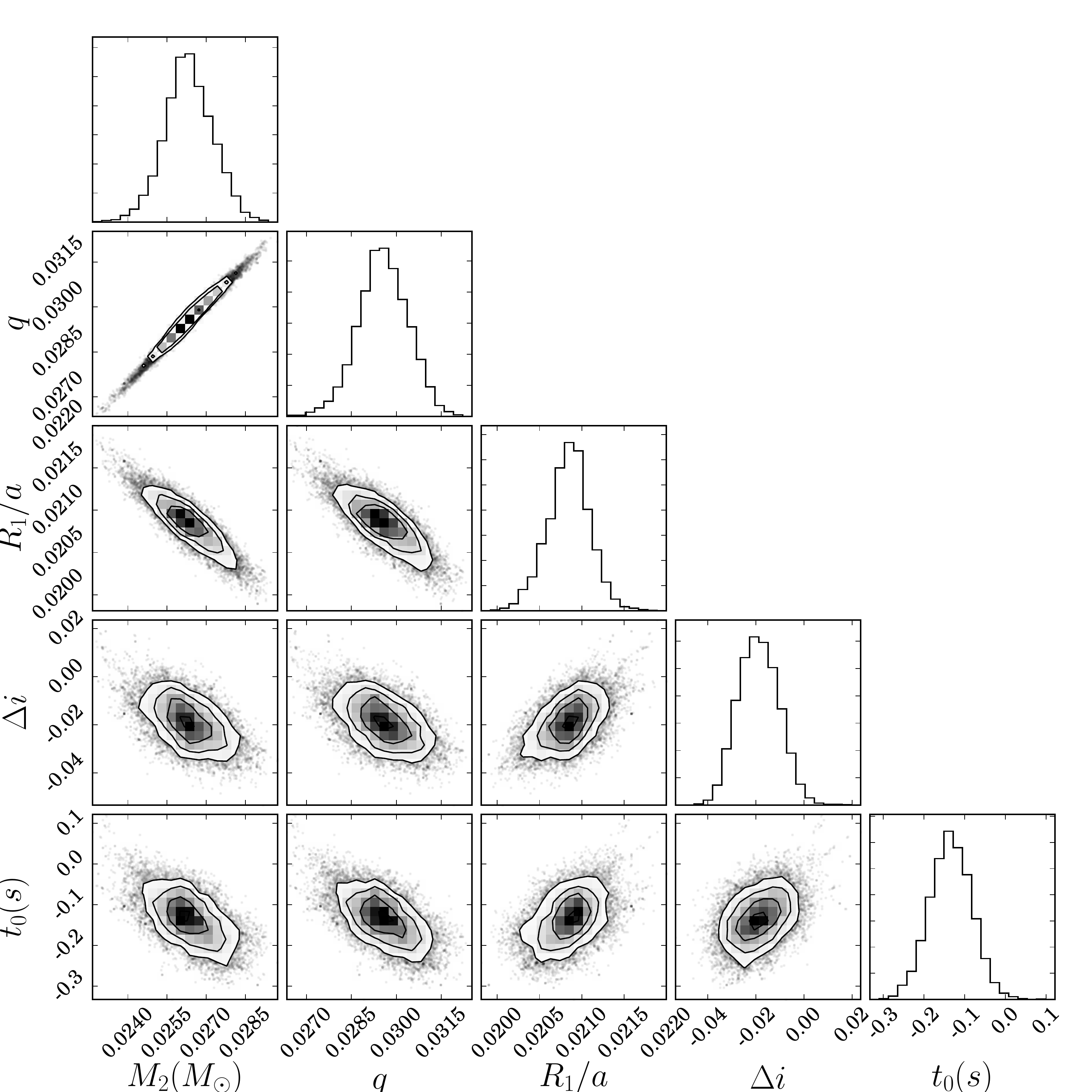}
    \caption{Corner plot of key parameters in the MCMC fit to the phase-folded {\fr} data. We also include $M_2$, which is not itself a parameter of the fit but is derived from $q$ and $R_1$.
    }
    \label{fig:corner-plot}
\end{figure}

The results of fits to individual eclipses are summarised in Table~\ref{tab:results}. The best-fit models to the phase-folded {\fg} and {\fr} lightcurves are shown in Figure~\ref{fig:averaged-eclipses}, and their parameters presented in Table~\ref{tab:av-results}. We show as an example the corner-plot of key fit parameters to the folded {\fr} data in Figure~\ref{fig:corner-plot}. 

Table~\ref{tab:av-results} also includes several observables that describe the eclipses: the phase width of the white dwarf eclipse from mid-ingress to mid-egress ($\Delta \phi$), the duration of the white dwarf ingress or egress ($\Delta w$), and the ingress and egress phases for the bright spot eclipse ($\phi_{\text{spot}, i}$,~$\phi_{\text{spot}, e}$). These parameters are calculated from the best fit models.

\begin{figure}
	\includegraphics[width=\columnwidth]{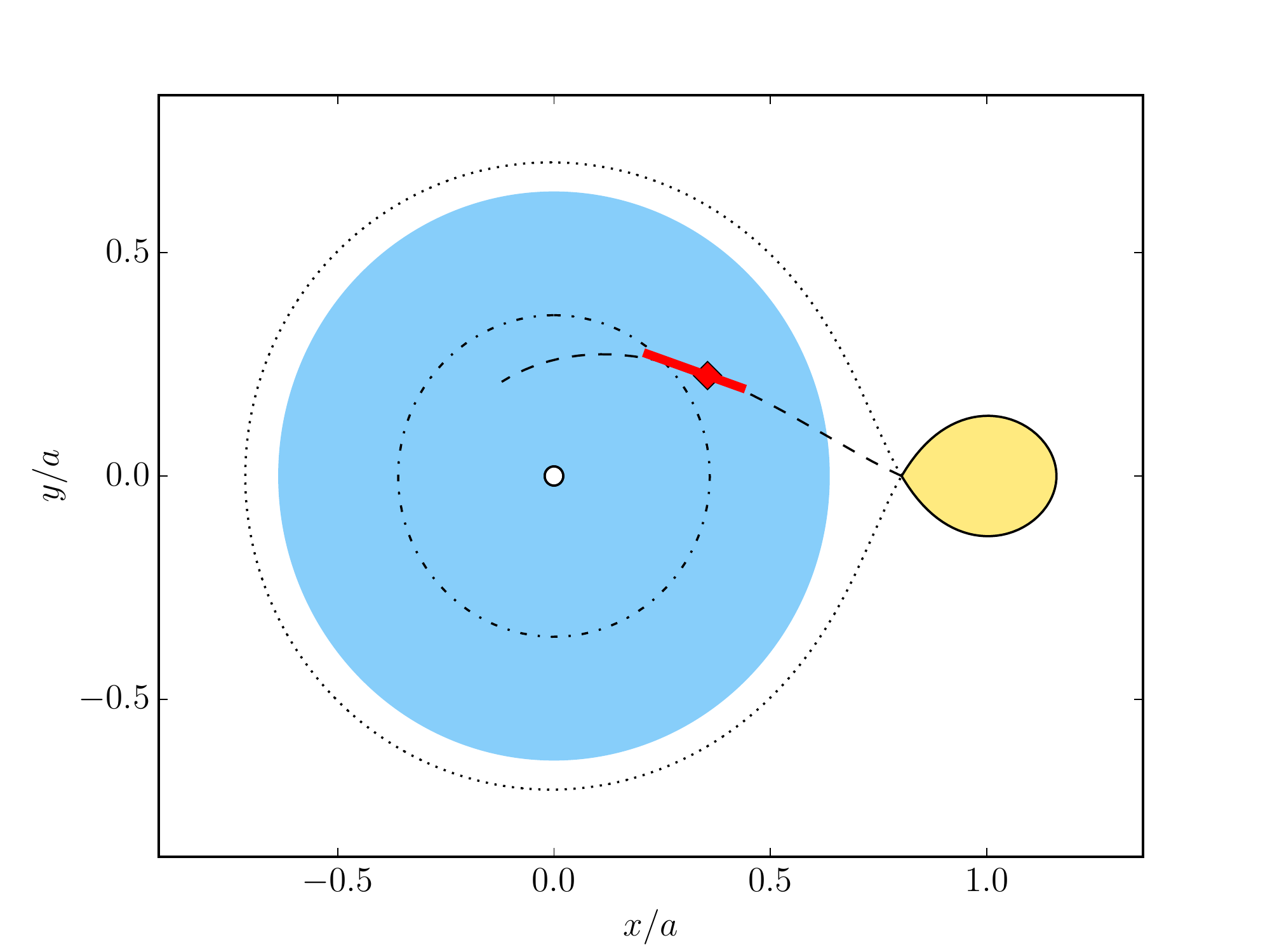}
    \caption{Schematic geometry of Gaia14aae calculated from the best-fit model to the phase-folded {\fr} data, showing the central white dwarf (white), the accretion disc (blue) and the donor star (yellow). The position of the bright spot peak luminosity is shown by the red diamond, with the red line showing the extension to half maximum of the bright spot. The path of the infalling stream is shown as a dashed line, the edge of the primary Roche lobe as a dotted line, and the circularisation radius as a dashed-dotted line.
    }
    \label{fig:geometry}
\end{figure}

The geometry of the best-fit model for the {\fr} phase-folded data is shown in Figure~\ref{fig:geometry}. The geometry of the disc and bright spot are particularly worthy of note. We find an accretion disc radius very close to the Roche lobe radius, and larger than the theoretical tidal limit of $0.58 a$ \citep[][but note this approximation assumes no viscosity, and begins to deviate from the true limit for $q<0.03$]{Paczynski1977}. The bright spot is not located on the edge of the disc, but closer to the circularisation radius. However, this is dependent on the modelled brightness profile of the disc, which may not accurately represent the physical disc. The bright spot in our best fit model also appears to be significantly extended along the direction of the infalling stream.

\begin{table*}
	\centering
	\caption{Summary of the modelling results based on the phase-folded data. The means given in the final column here are produced using a weighting of 1/$\sigma^2$. Uncertainties are the formal MCMC uncertainties except where marked.
	}
	\label{tab:av-results}
	\begin{tabular}{lccc} 
		\hline
		Param.\ & \fr & \fg & Mean\\
		\hline
		$\Delta \phi$ & $0.0373 \pm 0.0005$ & $0.0373 \pm 0.0004$ & $0.0373 \pm 0.0003$\\
		$\Delta w$ & $0.00787 \pm 0.00014$ & $0.00800 \pm 0.00011$ & $0.00795 \pm 0.00009$\\
		$\phi_{\text{spot}, i}$ & $0.0220 \pm 0.0007$ & $0.0234 \pm 0.0005$ & $0.0229 \pm 0.0004$\\
		$\phi_{\text{spot}, e}$ & $0.0863 \pm 0.0004$ & $0.0845 \pm 0.0005$ & $0.0856 \pm 0.0003$\\
		\hline
		$q$ & $0.0283 \pm 0.0007$ & $0.0290 \pm 0.0006$ & $0.0287 \pm 0.0020$ *\\
		$i$ $(^\circ)$ & $86.40 \pm 0.12$ & $86.27 \pm 0.10$ & $86.3 \pm 0.3$ *\\
		$M_1$ ($M_\odot$) & $0.870 \pm 0.007$ & $0.872 \pm 0.007$ & $0.87 \pm 0.02$ *\\
		$M_2$ ($M_\odot$) & $0.0246 \pm 0.0008$ & $0.0253 \pm 0.0007$ & $0.0250 \pm 0.0013$ *\\
		$R_1/a$ & $0.0215 \pm 0.0002$ & $0.0215 \pm 0.0002$ & $0.0215 \pm 0.0006$ *\\
		$R_2/a$ & $0.1393 \pm 0.0011$ & $0.1404 \pm 0.0009$ & $0.140 \pm 0.002$ *\\
		$R_\text{disc}/a$ & $0.570 \pm 0.008$ & $0.640 \pm 0.006$ & $0.615 \pm 0.005$\\
		$R_\text{spot}/a$ & $0.442 \pm 0.004$ & $0.421 \pm 0.005$ & $0.434 \pm 0.003$\\
		$a$ ($R_\odot$) & $0.430 \pm 0.001$ & $0.430 \pm 0.001$ & $0.430 \pm 0.003$ *\\
	\hline
	\multicolumn{4}{p{9cm}}{* On the starred results in the final column, error bars are inflated to include the estimated systematic uncertainty in $q$, as discussed in Section~\ref{masses}.}\\
	\hline
	\end{tabular}
\end{table*}

\subsection{Ephemeris}

\begin{figure*}
	\includegraphics[width=500pt]{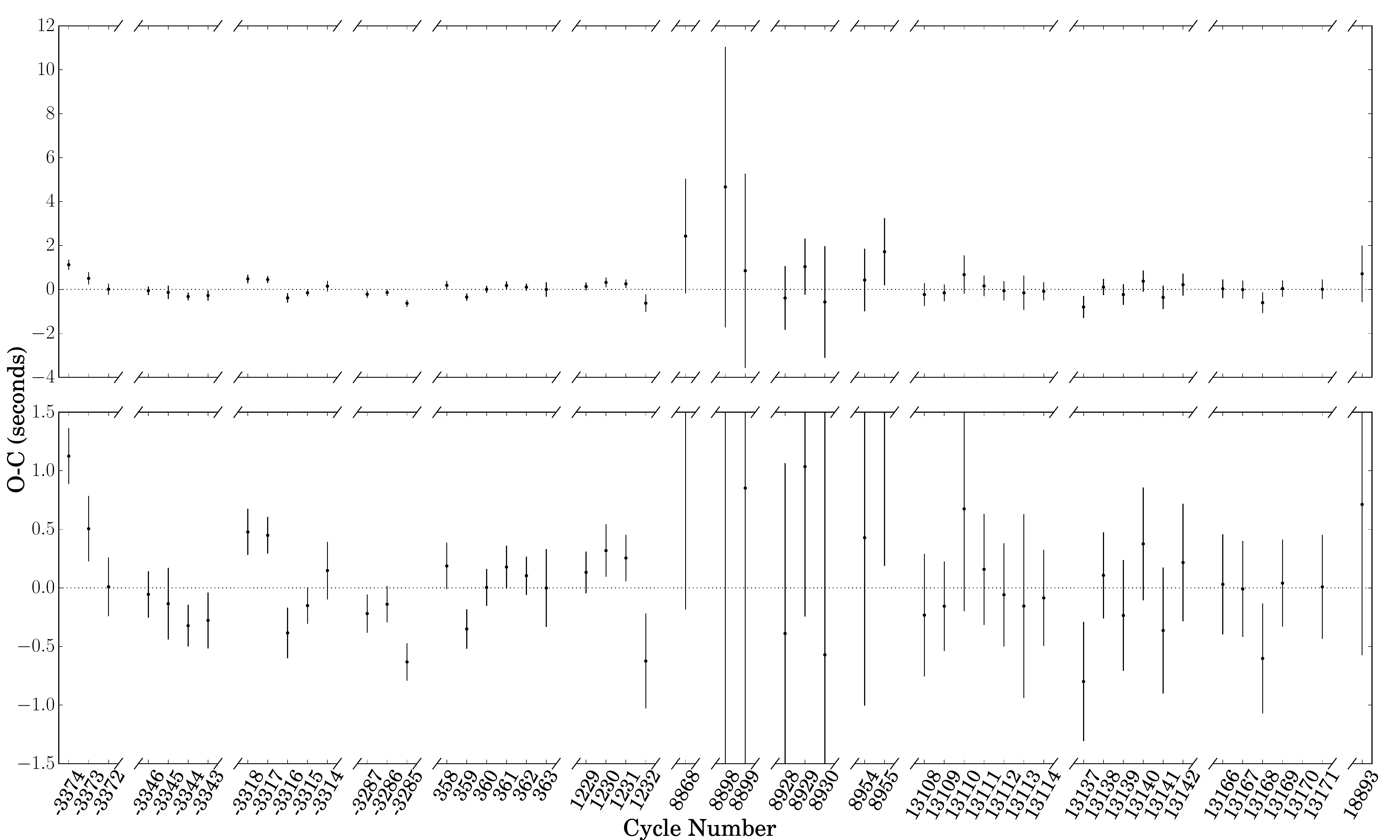}
    \caption{Mid-eclipse timing measurements of Gaia14aae, showing the difference between the measured times and those predicted from the ephemeris in Equation~\ref{eq:ephem}. The bottom panel is a zoomed-in version of the top. Measurements in multiple colour bands have been combined using a weighted mean for this diagram, although these were treated as separate eclipses for the calculation of the ephemeris. 
    }
    \label{fig:example_figure}
\end{figure*}

From the mid-eclipse times, $t_0$, for individual eclipses listed in Table~\ref{tab:results}, we calculated an ephemeris for Gaia14aae of
\begin{equation}
	\text{BMJD(TDB)} = 57153.6890966(4) + 0.03451957084(8) E
	\label{eq:ephem}
\end{equation}
where $E$ is the cycle number and the time of zero phase corresponds to the centre of the white dwarf eclipse. The quoted zero phase was chosen so as to minimise correlation between the zeropoint and period, and the quoted uncertainties are 1$\sigma$.

This ephemeris differs significantly from that quoted in \citet{Campbell2015}. This difference arises from an 11~s discrepancy in the BFOSC data used as part of that ephemeris calculation, possibly a systematic offset in the instrument which is not built for precise timings. Due to this and the relatively low time resolution of the lightcurves used by \citet{Campbell2015}, we did not include their data in our ephemeris calculation.

Including all data in individual colour bands we have 119 measured eclipse times, giving us 117 degrees of freedom. The $\chi^2$ of this linear ephemeris is 105.2.

A linear ephemeris agrees with expectations. The period growth of this system is expected to be dominated by angular momentum loss due to gravitational wave radiation. Using the \citet{Landau1971} formula
\begin{equation}
	\left(\frac{\dot{J}}{J}\right) = - \frac{32}{5} \frac{G^3}{c^5} \frac{M_1 M_2 (M_1+M_2)}{a^4}
\end{equation}
we would expect $\dot{J}/J = -1.6 {\times} 10^{-17} \text{~s$^{-1}$}$. \review{Under the assumptions that mass transfer is conservative and that angular momentum loss is completely dominated by gravitational wave radiation,}
$\dot{P}_\text{orb}$ can then be found \citep{Deloye2007} by
\begin{equation}
\label{gwradiation}
	\left(\frac{\dot{P}_{\text{orb}}}{P_{\text{orb}}}\right) = 3 \left(\frac{\dot{J}}{J}\right) \left[\frac{\xi_{R_2} - 1/3}{\xi_{R_2}+5/3-2q}\right]
\end{equation}
where $\xi_{R_2} = d(\log{R_2}) / d(\log{M_2})$. Taking $\xi_{R_2} \approx -0.2$ as the approximate gradient of the $M$--$R$ tracks in Figure~\ref{fig:compare-models} gives $\dot{P}_\text{orb} \approx 1.7{\times}10^{-6}$~s~yr$^{-1}$. Given these data, a quadratic term added to the ephemeris in Equation~\ref{eq:ephem} would not be detected to 3$\sigma$ unless its value were $\gtrsim 9{\times}10^{-6}$~s~yr$^{-1}$. The predicted change is therefore not detectable with these data. However, as the detectability scales with $t^2$, and our current baseline is only two years, we can expect the period change to become detectable within the next few years.

\citet{Copperwheat2011} found some evidence of departures from a linear ephemeris in the other eclipsing AM\,CVn type binary, YZ\,LMi, which were ascribed to systematic errors induced by flickering or the superhump period. To search for similar systematic biases in Gaia14aae, we checked for a correlation between the central times of each eclipse and the brightness of the disc as measured during that eclipse. We find no strong correlation, but there are two eclipses which have both an unusually bright disc and an unusually late eclipse time. Both are {\fili} measurements, out of only three eclipses which were measured in {\fili}. This is the band in which the disc is brightest. It is therefore possible that the {\fili} data may be biased in some way by the disc eclipse. In {\fu}, {\fg} and {\fr} we find no evidence for a similar correlation. 

We searched for significant periods in a Lomb-Scargle periodogram \citep{Lomb1976,Scargle1982} of these data outside of eclipse. Other than the orbital period and its harmonics, we found no significant periods. By comparing with injected, sinusoidal signals, we estimate that we would detect a signal of approximately 1\% strength or greater. After subtracting a sinusoid of period equal to the orbital period from the data and creating another Lomb-Scargle transform from the residuals, we still do not detect any other periods. Note that, given the white dwarf temperature of 12900K \citep{Campbell2015}, we do not expect to see DBV pulsations.

\subsection{White Dwarf Luminosity and Colour}
\label{luminosity}

\begin{figure}
	\includegraphics[width=\columnwidth]{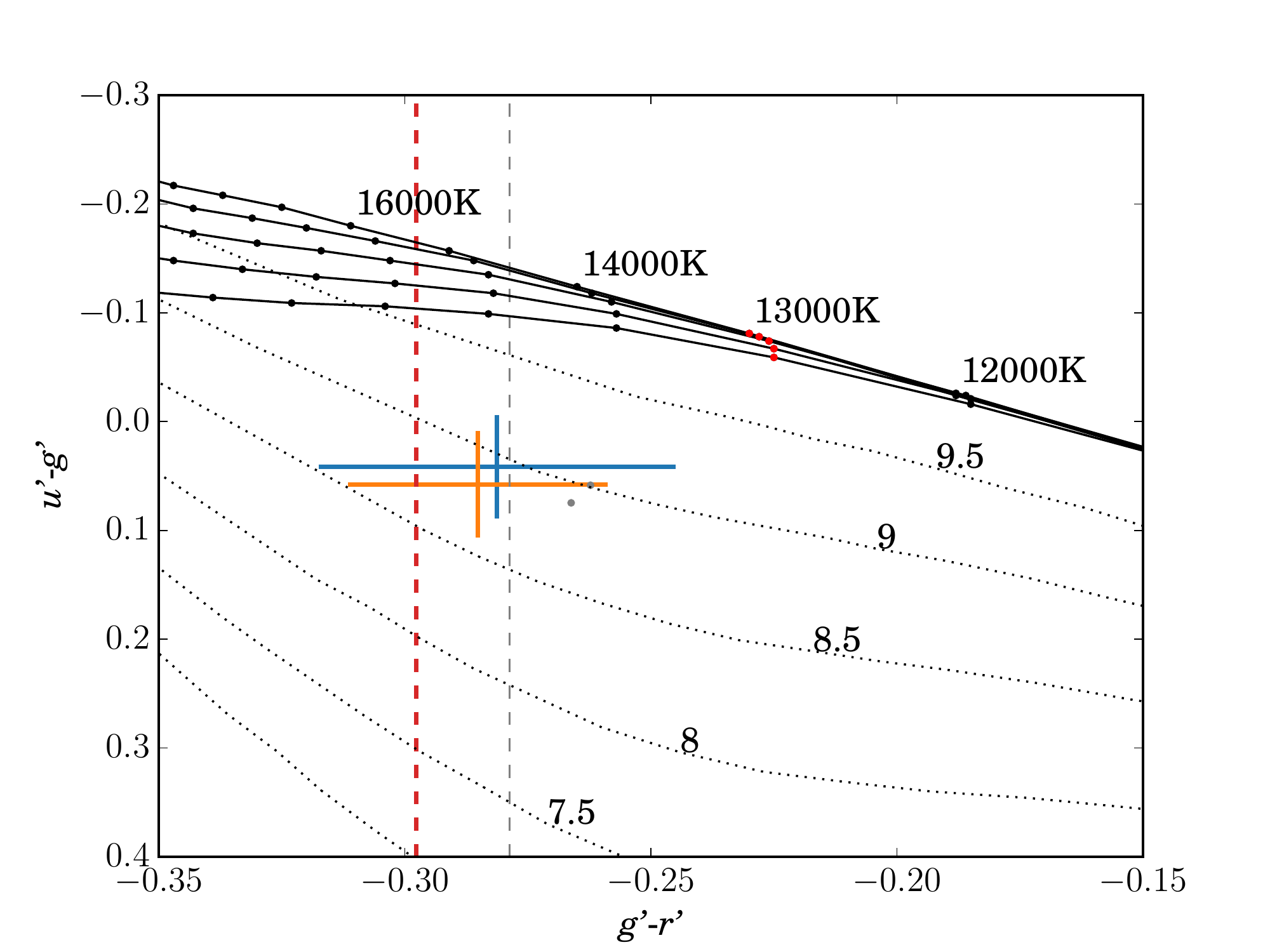}
    \caption{The position of the primary white dwarf in colour-colour space, as measured in January 2015 (cyan), May/June 2015 (orange), and August 2016 (red dashed line, {\fg}-{\fr} constraint only). These fluxes have been corrected for reddening. Grey points show the positions of these measurements prior to reddening correction. Also shown are DB model atmospheres (red dots 13000K models, black dots other temperatures, solid lines connecting DB models of constant surface gravity) and DA model atmospheres (dotted lines for models of constant surface gravity), with temperatures and surface gravities labelled \citep{Holberg2006,Kowalski2006,Tremblay2011,Bergeron2011}.
    }
    \label{fig:colours}
\end{figure}

\begin{figure}
	\includegraphics[width=\columnwidth]{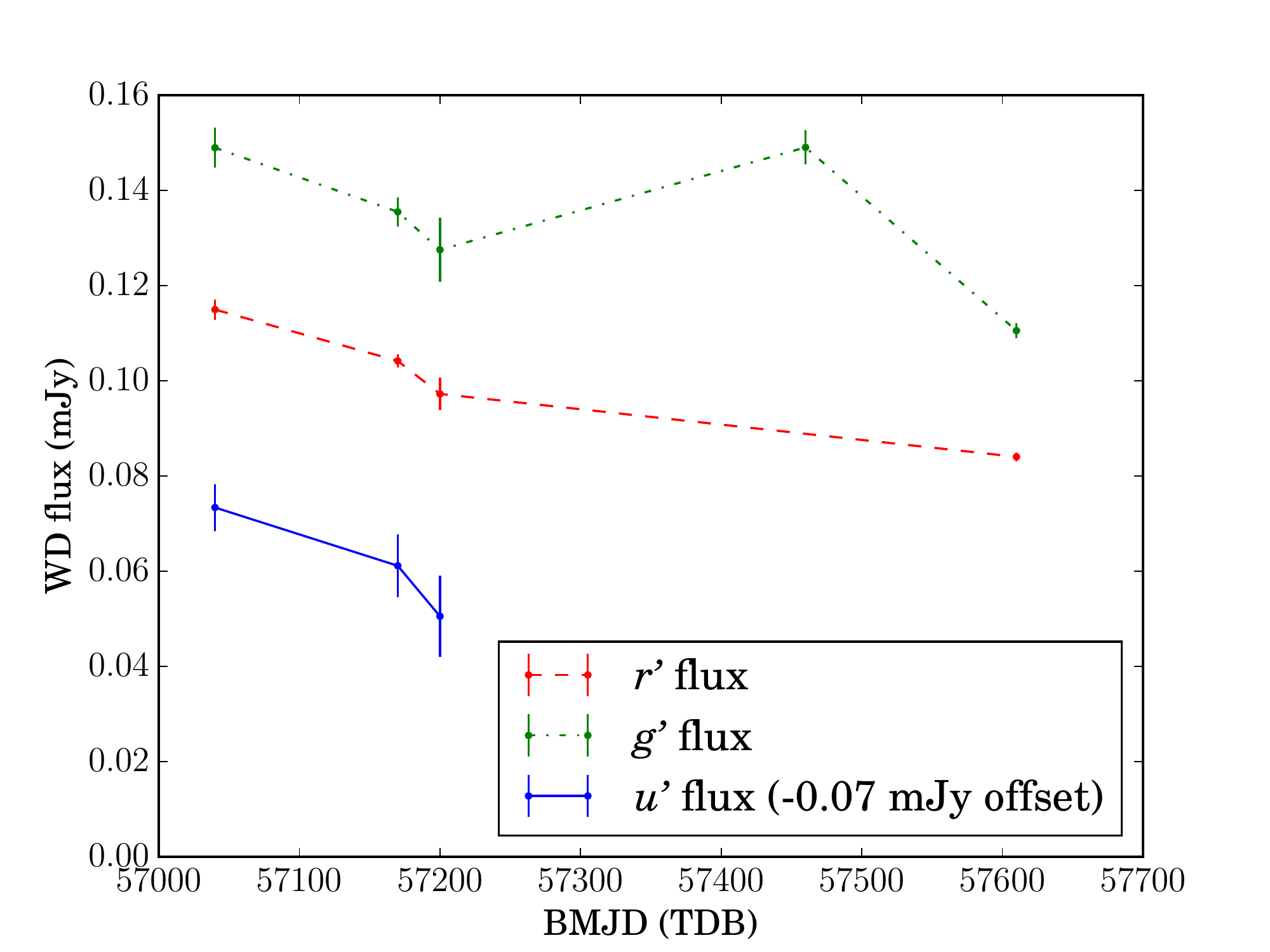}
    \caption{Evolution of the measured white dwarf flux over time, averaged between eclipse depth measurements from individual eclipses. These fluxes have been corrected for reddening. The {\fg} measurement from March 2016 (MJD $\sim$ 57500) is possibly biased by the large amount of flickering during these observations. The most recent outburst for this system occurred prior to the start of observations, at MJD = 56880.
    }
    \label{fig:luminosities}
\end{figure}

\begin{figure*}
	\includegraphics[width=500pt]{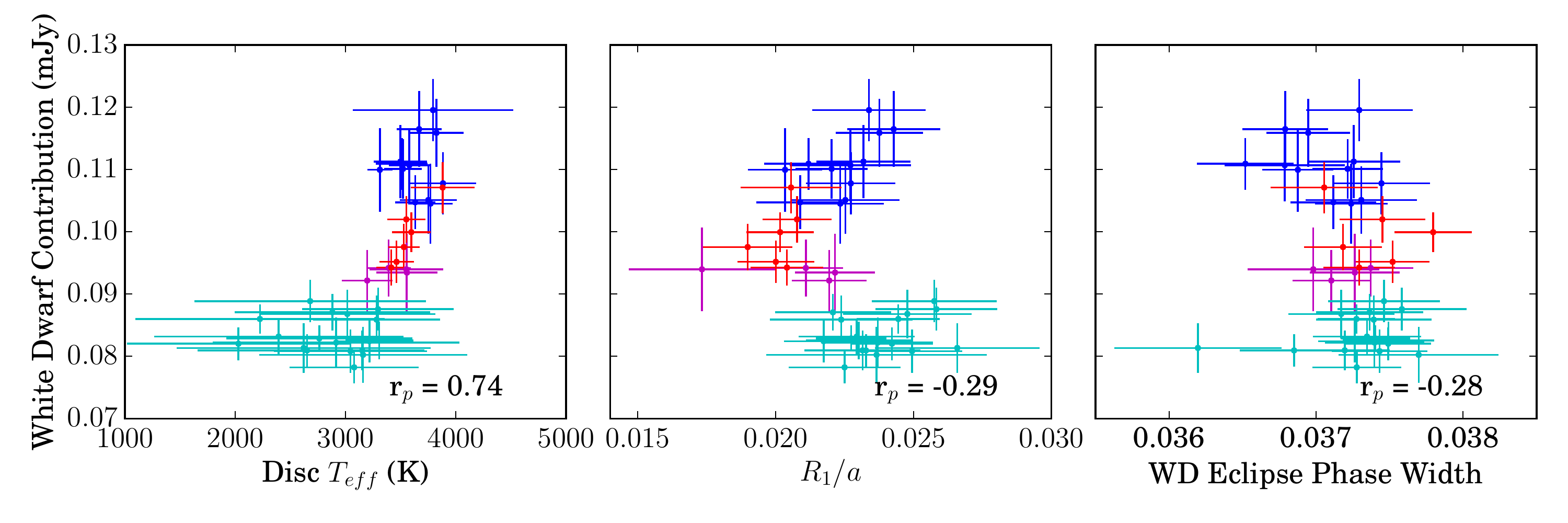}
    \caption{White dwarf luminosity as a function of other fit parameters across all ULTRACAM and CHIMERA {\fr}-band eclipses. The correlation coefficient quoted is Pearson's r. Colours represent the dates on which the data were taken: January 2015 (blue), May 2015 (red), June 2015 (magenta), and August 2016 (cyan).
    }
    \label{fig:correlations}
\end{figure*}

For each eclipse, we measured the flux contribution from the white dwarf in our best-fit model and calculated the \textit{u'-g'} and \textit{g'-r'} colour indices. We calculated these separately for January and May/June 2015, as well as the \textit{g'-r'} colour from August 2016. These are shown in Figure~\ref{fig:colours}. These colours are corrected for interstellar extinction according to \citet{Schlegel1998} and \citet{Schlafly2011}, using E(B-V)=0.018. We also show for comparison the expected colours from DB and DA white dwarf atmosphere models \citep{Holberg2006,Kowalski2006,Tremblay2011,Bergeron2011}. Our colours are approximately $2\sigma$ different from those expected for any DB white dwarf, although the closest model at 13000K is similar to the expected temperature of 12900K, based on its GALEX UV flux \citep{Campbell2015}. It should be noted that the white dwarf may not have a pure DB atmosphere; the accreted material may have a significant fraction of heavier elements, particularly C, N, O, and Ne, depending on the prior evolution of the donor \citep{Yungelson2008,Nelemans2010}.
The white dwarf lies in a region of colour space that is occupied by some DA white dwarfs, but its spectrum shows no sign of hydrogen and a DA white dwarf of this colour would require a higher temperature. If the white dwarf eclipse depths are indeed biased by some region of the disc as discussed below, this could be evidence of a colour dependence to that bias.

By comparing the {\fg} flux of the central white dwarf to that predicted for a DB white dwarf with a temperature of 13000K and log($g$) of 8.5 \citep{Holberg2006,Bergeron2011}, we estimate a distance of $188 \pm 13$pc. Of course, if there is some obscuration of the white dwarf by the disc, this estimate may not be reliable. The parallax that will be measured by \textit{Gaia} will provide a more reliable estimate of the distance to this system.

The flux of the white dwarf over time is shown in Figure~\ref{fig:luminosities}. Between January 2015 and August 2016, we measured a decrease of ($26 \pm 2$)\% in both {\fr} and {\fg} bands. The ratio between {\fr} and {\fg} stayed approximately constant.

This could be due to genuine cooling of the white dwarf, which would have been heated by its outbursts in 2014. Assuming the white dwarf is approximately described by a blackbody spectrum at 13000K, and treating the filters as top-hat functions whose value is 1 inside their FWHM and 0 elsewhere, this flux decrease would imply a temperature change of $\gtrsim$~1000K over the 18 month time interval (beginning approximately 150 days after the most recent outburst). This change is slightly less than the cooling seen in the short-period CV WZ\,Sge over the period 150-700 days after its outburst \citep{Godon2004}, and less than seen in GW Lib over the period 3-4 years after its outburst \citep{Szkody2012}. Such a temperature change would not produce a significant colour change, which is consistent with our observations.

\review{It should be noted that the 12900K measurement of \citet{Campbell2015} is higher than the $\approx 11000$K temperature prediction of \citet{Bildsten2006}, which was based on compressional heating of the white dwarf by the accreted matter. \citet{Bildsten2006} predict that the temperature of the central white dwarf is driven by accretion-induced heating until the system reaches an orbital period of around 40~minutes. For periods longer than this the decreased mass accretion rate means that the temperature decouples from accretion-driven heating, and the white dwarf then follows standard cooling tracks. However, as will be discussed in the following sections, the donor star in Gaia14aae seems to have a higher mass and radius than was assumed by \citet{Bildsten2006}, resulting in a higher mass transfer rate, which may explain this elevated temperature.}

While a change in the temperature of the central white dwarf is one possible explanation of the change in eclipse depth, other possibilities must be considered. This change in eclipse depth could be explained if there were a component of the system which is not included in our models and which is not resolved from the white dwarf eclipse, such that the apparent dimming of the white dwarf is in fact the dimming of this other component. \citet{Wood1986a} and \citet{Spark2015} discussed the optical visibility of the `boundary layer' through which the white dwarf accretes from its disc. Though this boundary layer can theoretically be resolved from the white dwarf by the shape of the eclipse ingress and egress, doing so would require a higher S/N than is currently available for this system. If present, a bright equatorial boundary layer might bias measurements of the white dwarf radius towards smaller or larger values, and a variable boundary layer might therefore induce a correlation between the modelled white dwarf luminosity and radius. We checked for such a correlation in our best-fit models to individual eclipses, as shown in Figure~\ref{fig:correlations}. We did not find a significant correlation. 


We do find a reasonably strong correlation between the temperature of the white dwarf and that of the disc. This could be evidence that some disc luminosity is not resolved from the white dwarf eclipse, biasing our measurement of white dwarf luminosity. Alternatively, there may be some causal link between the two, if both are heated by accretion or if both are still decreasing following the 2014 outburst.

The measurement of $q$ described in the next section is constrained by the contact phases of white dwarf and bright spot eclipses. Therefore, $q$ should not be affected by a bias in the apparent depth of the white dwarf eclipse. However, if the same mechanism were to bias the phase width of the white dwarf eclipse this would be a problem. We therefore searched for a correlation between the depth and phase width of the white dwarf eclipse as measured from our model fits, but found no evidence for a significant correlation (see Figure~\ref{fig:correlations}).

\subsection{Mass Ratio and Donor Mass}
\label{masses}

\begin{table}
	\centering
	\caption{Summary of the $q$ values found by the different methods described in Section~\ref{Analysis}. 	}
	
	\label{tab:q-values}
	\begin{tabular}{lc} 
		\hline
		Method & $q$\\
		\hline
		Phase-folded lightcurves MCMC & $0.0287 \pm 0.0005$\\
		Individual lightcurves MCMC & $0.030 \pm 0.003$\\
		Bootstrapping method & $0.0281 \pm 0.0007$\\
		Contact phase measurement & $0.0267 \pm 0.0012$\\
	\hline
	\end{tabular}
\end{table}

From the phase width of the white dwarf eclipse, a lower limit on the mass ratio can be found by assuming $i=90^\circ$. The phase width shown in Table~\ref{tab:av-results} gives $q_{min} = 0.0185 \pm 0.0005$. This is comparable to the lower limit found by \citet{Campbell2015}, $q_{min} = 0.019$.

Thus far in this paper we have described several methods of determining $q$: MCMC fitting to phase-folded lightcurves and to lightcurves of individual eclipses, fitting to a collection of bootstrapped phase-folded lightcurves, and measurement of bright spot contact phases by hand. The results found by each of these methods are summarised in Table~\ref{tab:q-values}.

The most rigorous method for calculating $q$ is the phase-folded MCMC method, and hence we favour the result from this method. However, the scatter of the results in Table~\ref{tab:q-values} suggests that the formal uncertainty on $q$ quoted by the MCMC may be an underestimate. This is not unexpected, as the method does not take into account any systematic biases that may come from the variability of the source between eclipses. We therefore suggest a more conservative error bar of 0.0020, in an attempt to take this scatter into account. Our canonical mass ratio is then $q = 0.0287 \pm 0.0020$.

To find the corresponding uncertainties on other stellar parameters, most importantly $M_2$, we propagated this uncertainty through the derivation of the other parameters from the raw observables $\Delta \phi$ and $\Delta w$. The resulting uncertainties are shown in the marked entries in Table~\ref{tab:av-results}.

\subsection{Comparison to Models}

\begin{figure*}
	\includegraphics[width=500pt]{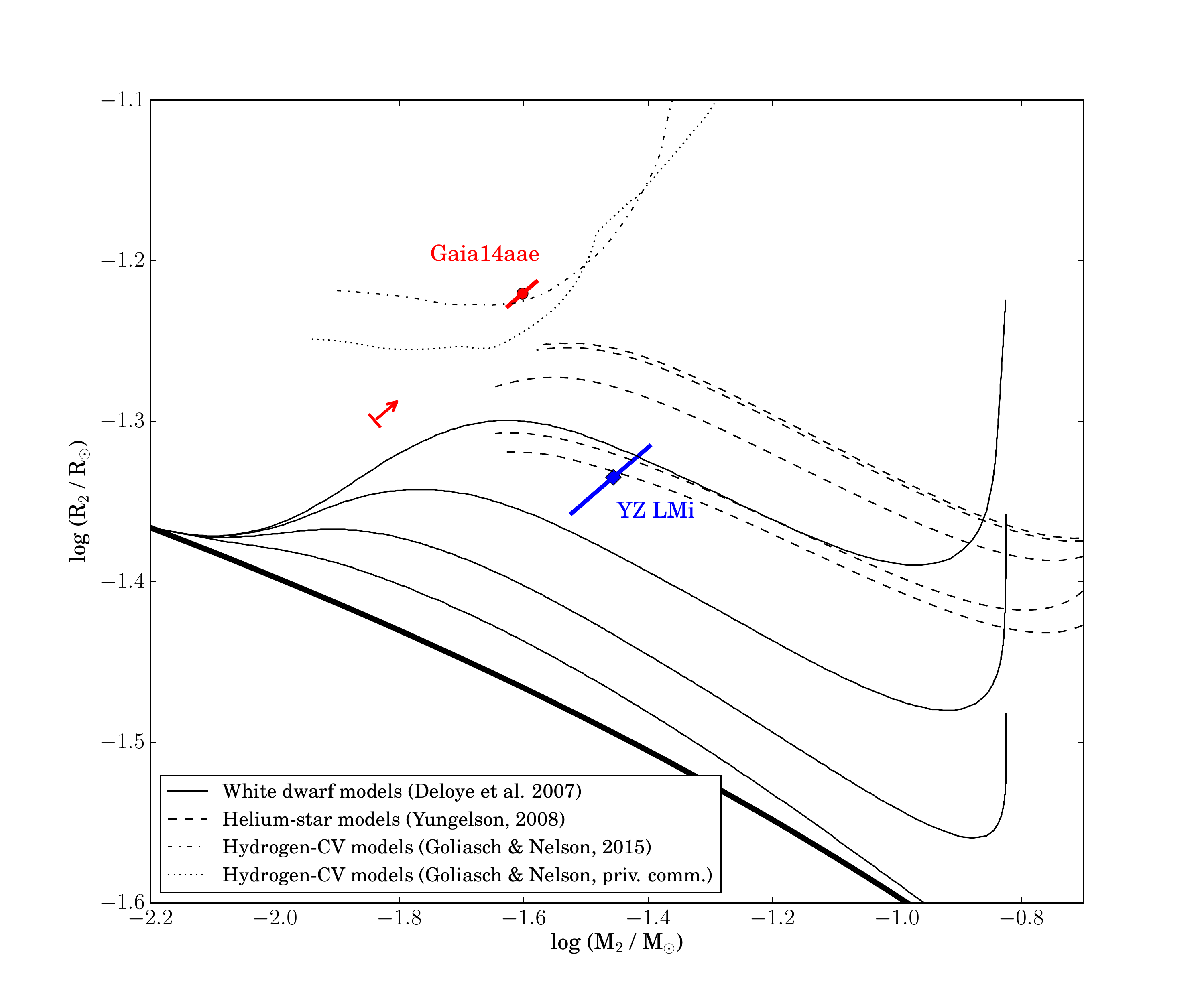}
    \caption{The measured donor masses and radii of Gaia14aae (red circle) and J0926 (blue diamond) and their uncertainties (solid lines of corresponding colours) compared to evolutionary tracks from \citet[thin black lines]{Deloye2007}, \citet[dashed]{Yungelson2008}, \citet[dashed-dotted]{Goliasch2015}, and Goliasch \& Nelson (priv.~comm., dotted). The white dwarf and helium star donor tracks include multiple levels of degeneracy, with the most degenerate models at the bottom. We also show the $M$--$R$ track for a zero-entropy donor (thick solid line). The arrow shows the lower limit on $M_2$ corresponding to $q_\text{min} = 0.0185$ for Gaia14aae. The diagonal uncertainties result from the strong constraints on the donors' mean densities resulting (according to Equation~\ref{eq:mr}) from their tightly constrained orbital periods.
    }
    \label{fig:compare-models}
\end{figure*}

The evolutionary history of Gaia14aae can be explored by comparing the mass of the donor star with theoretical evolutionary tracks. 
From Kepler's third law and the constraint that the radius of the donor star equals the radius of the Roche lobe comes the relation between donor mass and radius for a given orbital period \citep{Faulkner1972a}
\begin{equation}
P_\text{orb} = 101\text{s}\left(\frac{R_2}{0.01R_\odot}\right)^{\frac{3}{2}}\left(\frac{0.1M_\odot}{M_2}\right)^{\frac{1}{2}}.
\label{eq:mr}
\end{equation}
This can alternately be expressed as a constraint on mean donor density coming from the orbital period alone. Therefore, our measurement of $M_2$ combined with the orbital period of Gaia14aae ($P_{\text{orb}} = 49.71$~minutes) can locate the donor in $M$--$R$ space, allowing us to compare the donor properties with the evolutionary tracks.

We take numerically calculated $M$--$R$ evolution tracks for AM\,CVn donors that have been published for the white dwarf donor \citep{Deloye2007}, helium-star donor \citep{Yungelson2008}, and evolved-CV \citep{Goliasch2015} formation channels. We show these tracks in Figure~\ref{fig:compare-models}.

We show white dwarf donor models with total binary mass $M_\text{tot} = 0.825\text{M}_\odot$, similar to our measured $M_\text{tot} \approx 0.91M_\odot$ \citep{Deloye2007}. These tracks are for a range of initial donor degeneracies, with the top tracks being the least degenerate. The highest of these denotes the approximate boundary between the least degenerate white dwarf donor tracks and the most degenerate helium-star donor tracks. The helium-star donor tracks are for a binary with initial conditions $M_{1,i} = 0.8M_\odot$, $M_{2,i} = 0.65M_\odot$ and are distinguished by the evolutionary status of the donor, with the top tracks being the least evolved and the bottom tracks being the most evolved \citep{Yungelson2008}.

For the white dwarf and helium-star evolution tracks, a donor would evolve into contact at the right-hand side of Figure~\ref{fig:compare-models} and evolve through mass transfer to a lower mass and longer orbital period (moving right-to-left in Figure~\ref{fig:compare-models}). At a period of $\approx$ 40 minutes, both \citet{Deloye2007} and \citet{Yungelson2008} predict that the thermal timescale of the donor becomes shorter than the mass-loss timescale, allowing the donor to cool in response to mass loss. The donor therefore loses entropy and contracts towards complete degeneracy. For both evolutionary channels, donors with orbital periods significantly longer than this are expected to be completely degenerate.

Our measured mass and radius lie outside the predicted parameter space for the white dwarf donor evolutionary channel. Given the density constraint arising from its period, the true donor mass would have to be a factor of 2-3 times smaller to agree with the tracks shown here. Indeed, the density constraint coupled with the minimum donor mass found by \citet{Campbell2015} are sufficient to disagree with the predicted tracks, including the collapse due to cooling, without needing the mass estimate presented in this work. Irradiation of the donor can delay the predicted collapse to longer periods. Even so, our mass estimate would lie significantly above the least degenerate white dwarf donor track. We therefore find that Gaia14aae is unlikely to have evolved through the white dwarf donor track.

The helium-star channel evolutionary tracks calculated by \citet{Yungelson2008} end once the AM\,CVns reach periods of around 40 minutes, making it difficult to compare our measured mass and radius to these tracks. The predicted collapse towards degeneracy beyond this point (the beginning of which is just visible at the left-hand end of these tracks) would disagree with our measurements. However, if this collapse were delayed to a period of 50 minutes or longer, our measurements would appear to be close to agreement with the least degenerate of the tracks shown here. Delaying this collapse to longer periods may be possible by, for instance, increasing the efficiency of irradiation of the donor by the accretor \citep{Deloye2007} or by outbursts.

We also show two evolved-CV tracks. The first, from \citet[][dashed-dotted line]{Goliasch2015}, has a 0.6~$M_\odot$ accretor, $\dot{M} = 1.4 {\times} 10^{-11} M_\odot \text{yr}^{-1}$, and a hydrogen fraction of $\sim$ 5\% in the transferred gas. The second track (Goliasch \& Nelson, priv. comm., dotted line) has an accretor of $0.85M_\odot$. These tracks are roughly representative of the edge of region of $M$--$R$ space this channel explores; systems above and to the left of these tracks (including Gaia14aae) could easily have formed by this channel. 
However, AM\,CVns that form by this channel are expected to retain a greater fraction of their hydrogen than AM\,CVns that form by the other two channels. Evolved-CV tracks with a smaller hydrogen fraction than $\sim 1\%$ in this region of $M$--$R$ space are possible, but they can only form from progenitor CVs with a limited range of initial parameters, and require long timescales to evolve. We might therefore expect to see a number of short-period CVs with visible hydrogen and helium for each AM\,CVn that forms by this channel. At present few such CVs are known \citep[see][for examples]{Augusteijn1996,Breedt2012}.



In terms of its measured mass and radius, the donor in Gaia14aae appears to fit best with tracks of the evolved-CV channel. However, the predicted hydrogen content of systems formed by these tracks is not seen in spectra of Gaia14aae \citep{Campbell2015}. \review{Though no quantitative upper limit exists for Gaia14aae, a hydrogen abundance on the order predicted for this channel would be easily detectable: modelling of the AM\,CVn systems GP\,Com \citep[with local thermal equilibrium models,][]{Marsh1991} and V396\,Hya \citep[with non-local thermal equilibrium models,][]{Nagel2009} find that any hydrogen abundances of $\gtrsim 10^{-5}$ should result in detectable Balmer emission. Similar upper limits have been found for DB white dwarfs in this temperature range \citep{Koester2015,Bergeron2011}.}
Producing similar tracks without detectable hydrogen would require finely-tuned starting parameters that make them unlikely (Goliasch \& Nelson, priv.~comm.). The absence of any visible hydrogen in the optical spectrum of Gaia14aae is therefore difficult to reconcile with predictions for this channel.

We conclude that Gaia14aae may have formed by either the evolved-CV channel or the helium-star donor channel. The results presented here are not in perfect agreement with either of the models as they currently stand, but one or both of these channels may have a region of parameter space that can explain the properties of Gaia14aae. The close agreement of our measured $M_2$ with that of the evolved-CV channel is particularly intriguing, given that this channel has generally been considered the least probable formation channel. The clear disagreement of our result with the white dwarf donor formation channel, which has been widely considered to be the dominant channel, is also of interest in light of the prediction by \citet{Shen2015a} that double white dwarf binaries will not reach a state of stable accretion.

\subsection{Implications For Gravitational Wave Emission}

The distribution of AM\,CVn-originated gravitational wave strains detectable by the Laser Interferometer Space Antenna (LISA) depends on the distribution of both periods and stellar masses across AM\,CVns. The donor mass of Gaia14aae presented above is greater than would be expected for a fully-degenerate donor in a binary at this orbital period. Consequently, the gravitational wave emission will be greater. 

Due to the high inclination of Gaia14aae, the $h_{+}$ component of its gravitational wave emission will dominate over the $h_{\times}$ component. The amplitude of the $h_{+}$ component can be calculated \citep{Korol2017} by 
\begin{equation}
h_{+} = \frac{2(G\mathcal{M}) ^ {5/3} (2 \pi / P_\text{orb})^{2/3}}{c^4d}(1+\cos^2i)
\end{equation}
where $\mathcal{M} = (M_1 M_2)^{3/5} (M_1 + M_2)^{-1/5}$ is the chirp mass of the system and $d$ is its distance from the Earth. 
For the stellar masses shown above, we predict the strain from Gaia14aae to be 
\begin{equation}
h_{+} = 1.2{\times}10^{-20} \times \frac{1\text{pc}}{d}.
\end{equation}
For the distance of 188pc we estimated in Section~\ref{luminosity}, this would give a strain of $6.1{\times}10^{-23}$. This is below the detection limit of LISA, unsurprisingly given the long orbital period of Gaia14aae. AM\,CVns at the short end of their period distribution are expected to be the brightest emitters of gravitational waves.

For comparison we consider a degenerate donor with the same orbital period but a mass of $M_2 = 0.01 M_\odot$. The strain for that system would be $h_{+} = 2.3{\times}10^{-23}$.
The gravitational wave emission of Gaia14aae is therefore a factor of 2.7 higher than would be expected in the degenerate case, and the volume of space in which the system would be detectable is increased by nearly a factor of 20.
This emphasises the need to understand the nature of AM\,CVn donors in order to predict the distribution of strains LISA will detect, both as resolved sources and as unresolved background. The results from both Gaia14aae and YZ\,LMi suggest that non-degenerate and partially degenerate donors are more common among AM\,CVns than previously believed. If this is the case, we would expect AM\,CVns as a population to be brighter emitters of gravitational waves for a given orbital period than previously predicted.

\section{Conclusions}
\label{Conclusions}

Gaia14aae is the third known eclipsing AM\,CVn-type binary and the first in which the central white dwarf is fully eclipsed. As such, it provides an unprecedented opportunity to measure the properties of the component stars in one of these systems and thereby constrain the system's prior evolution. The results are difficult to reconcile with existing models. 

\review{Any measurement of the properties of Gaia14aae is complicated by several unfortunate peculiarities of the system. In particular, the weakness of a key feature of the system (its `bright spot') means that lightcurve fitting can be biased by the intrinsic red noise of the system. These difficulties increase the systematic uncertainty in our measurements. We have attempted to take this uncertainty into account by inflating the quoted error bars on our results. 
}

We measured a mass ratio $q = 0.0287 \pm 0.0020$ and a donor mass $M_2 = 0.0250 \pm 0.0013 M_\odot$.  Combined with the donor density constraint arising from the orbital period, this mass shows that the donor is not degenerate and that the system did not evolve from a double degenerate binary. The system therefore must either have a non-degenerate helium star as its donor or be descended from a hydrogen CV with an evolved donor. In both cases there are unexplained questions: the donor in the former is expected to have collapsed towards degeneracy before reaching this orbital period, and the latter is expected to show traces of spectroscopic hydrogen. Neither of these predictions are observed, but it may be possible to tweak the models in order to explain Gaia14aae's evolution.


\section*{Acknowledgements}

The authors would like to thank Lorne Nelson and Lev Yungelson for exceedingly helpful discussions and insight, \review{and the anonymous referee for their constructive feedback}. 

MJG acknowledges funding from an STFC studentship via grant ST/N504506/1. TRM, DTHS and EB acknowledge STFC via grants ST/L000733/1 and ST/P000495/1. SB acknowledges funding from NWO VIDI grant 639.042.218, financed by the Netherlands Organisation for Scientific Research (NWO). VSD, SPL, ULTRACAM and ULTRASPEC are funded by STFC via consolidated grant ST/J001589. Support for this work was provided by NASA through Hubble Fellowship grant \#HST-HF2-51357.001-A, awarded by the Space Telescope Science Institute, which is operated by the Association of Universities for Research in Astronomy, Incorporated, under NASA contract NAS5-26555. Part of the research was carried out at the Jet Propulsion Laboratory, California Institute of Technology, under a contract with the National Aeronautics and Space Administration.

This publication made use of the packages \texttt{lcurve} \citep{Copperwheat2010}, \texttt{numpy}, \texttt{matplotlib}, \texttt{astropy}, \texttt{scipy}, \texttt{emcee} \citep{Foreman-Mackey2013}, and \texttt{corner} \citep{Foreman-Mackey2016}. DA and DB model white dwarf atmospheres for Figure~\ref{fig:colours} were taken from http://www.astro.umontreal.ca/{\textasciitilde}bergeron/CoolingModels.

The data presented in this work were obtained using the William Herschel Telescope (WHT) operated by the Isaac Newton Group (ING) at the Roque de los Muchachos Observatory on La Palma, the 2.4~m Thai National Telescope (TNT) operated by the National Astronomy Research Institute of Thailand (NARIT) at the Thai National Observatory on Doi Inthanon, and the 200-inch Hale Telescope at Palomar Observatory operated by the California Institute of Technology.




\bibliographystyle{mnras}
\bibliography{refs} 




\appendix

\section{Results of Individual Eclipse Modelling}

Table~\ref{tab:results} shows key parameters for the model fits carried out separately on individual eclipses.

\begin{table*}
	\centering
	\caption{Results from the fits carried out to individual eclipses. $\sigma$ here is the standard deviation of MCMC chain values. We also quote the weighted mean (with its propagated error) and standard deviation of these values for each parameter.}
	\label{tab:results}
\begin{tabular}[c]{l c c c c c c c c c c}
\hline
Observation date and filter & $q$ & $\sigma_q$ & $M_1 (M_\odot)$ & $\sigma_{M_1}$ & $M_2 (M_\odot)$ & $\sigma_{M_2}$ & $R_1 / a$ & $\sigma_{R_1/a}$ & $t_0$ (BTDB(MJD)) & $\sigma_{t_0}$\\
\hline
14 Jan 2015, Eclipse 1, \textit{u'} & 0.048 & 0.058 & 0.69 & 0.296 & 0.041 & 0.069 & 0.032 & 0.017 & 57037.2200876 & 2.08e-05 \\
14 Jan 2015, Eclipse 2, \textit{u'} & 0.047 & 0.06 & 0.777 & 0.25 & 0.046 & 0.079 & 0.026 & 0.01 & 57037.2545951 & 1.49e-05 \\
14 Jan 2015, Eclipse 3, \textit{u'} & 0.034 & 0.046 & 0.722 & 0.25 & 0.031 & 0.057 & 0.029 & 0.011 & 57037.289103 & 1.48e-05 \\
14 Jan 2015, Eclipse 1, \textit{g'} & 0.026 & 0.01 & 0.813 & 0.076 & 0.021 & 0.011 & 0.024 & 0.003 & 57037.2200773 & 2.8e-06 \\
14 Jan 2015, Eclipse 2, \textit{g'} & 0.032 & 0.018 & 0.865 & 0.098 & 0.029 & 0.022 & 0.022 & 0.003 & 57037.254589 & 3.5e-06 \\
14 Jan 2015, Eclipse 3, \textit{g'} & 0.029 & 0.015 & 0.824 & 0.085 & 0.025 & 0.018 & 0.023 & 0.003 & 57037.2891028 & 2.9e-06 \\
14 Jan 2015, Eclipse 1, \textit{i'} & 0.032 & 0.02 & 0.846 & 0.162 & 0.028 & 0.024 & 0.023 & 0.006 & 57037.2200826 & 1.56e-05 \\
14 Jan 2015, Eclipse 2, \textit{i'} & 0.058 & 0.077 & 0.69 & 0.194 & 0.052 & 0.086 & 0.03 & 0.009 & 57037.2546072 & 1.36e-05 \\
14 Jan 2015, Eclipse 3, \textit{i'} & 0.114 & 0.229 & 0.963 & 0.187 & 0.134 & 0.301 & 0.019 & 0.006 & 57037.2891138 & 9.4e-06 \\
15 Jan 2015, Eclipse 4, \textit{r'} & 0.046 & 0.041 & 0.929 & 0.139 & 0.047 & 0.051 & 0.02 & 0.005 & 57038.2901674 & 5.5e-06 \\
15 Jan 2015, Eclipse 3, \textit{r'} & 0.037 & 0.029 & 0.86 & 0.111 & 0.035 & 0.035 & 0.022 & 0.004 & 57038.2556489 & 4e-06 \\
15 Jan 2015, Eclipse 2, \textit{r'} & 0.044 & 0.024 & 0.901 & 0.106 & 0.042 & 0.028 & 0.021 & 0.004 & 57038.2211316 & 9e-06 \\
15 Jan 2015, Eclipse 1, \textit{r'} & 0.032 & 0.017 & 0.961 & 0.098 & 0.032 & 0.021 & 0.019 & 0.003 & 57038.1866147 & 4.6e-06 \\
15 Jan 2015, Eclipse 4, \textit{g'} & 0.036 & 0.025 & 0.913 & 0.112 & 0.035 & 0.031 & 0.02 & 0.004 & 57038.2901686 & 3e-06 \\
15 Jan 2015, Eclipse 3, \textit{g'} & 0.024 & 0.006 & 0.829 & 0.058 & 0.02 & 0.007 & 0.023 & 0.002 & 57038.255648 & 2.3e-06 \\
15 Jan 2015, Eclipse 1, \textit{g'} & 0.026 & 0.009 & 0.881 & 0.067 & 0.023 & 0.011 & 0.021 & 0.002 & 57038.186611 & 2.6e-06 \\
15 Jan 2015, Eclipse 4, \textit{u'} & 0.025 & 0.018 & 0.715 & 0.199 & 0.019 & 0.019 & 0.029 & 0.009 & 57038.2901634 & 1.84e-05 \\
15 Jan 2015, Eclipse 3, \textit{u'} & 0.024 & 0.015 & 0.698 & 0.191 & 0.018 & 0.018 & 0.029 & 0.008 & 57038.2556491 & 1.33e-05 \\
15 Jan 2015, Eclipse 2, \textit{u'} & 0.037 & 0.038 & 0.707 & 0.205 & 0.031 & 0.042 & 0.029 & 0.009 & 57038.2211398 & 1.77e-05 \\
15 Jan 2015, Eclipse 1, \textit{u'} & 0.038 & 0.032 & 0.77 & 0.263 & 0.035 & 0.042 & 0.027 & 0.012 & 57038.1866068 & 1.63e-05 \\
15 Jan 2015, Eclipse 2, \textit{g'} & 0.04 & 0.021 & 0.935 & 0.091 & 0.039 & 0.025 & 0.019 & 0.003 & 57038.2211279 & 3.1e-06 \\
16 Jan 2015, Eclipse 5, \textit{g'} & 0.029 & 0.013 & 0.833 & 0.086 & 0.025 & 0.015 & 0.023 & 0.003 & 57039.2912408 & 3.3e-06 \\
16 Jan 2015, Eclipse 5, \textit{r'} & 0.053 & 0.039 & 0.912 & 0.137 & 0.053 & 0.047 & 0.02 & 0.005 & 57039.2912385 & 6.6e-06 \\
16 Jan 2015, Eclipse 4, \textit{r'} & 0.031 & 0.014 & 0.866 & 0.092 & 0.028 & 0.017 & 0.022 & 0.003 & 57039.2567161 & 3.5e-06 \\
16 Jan 2015, Eclipse 3, \textit{r'} & 0.027 & 0.007 & 0.791 & 0.097 & 0.021 & 0.008 & 0.025 & 0.004 & 57039.2221936 & 5.2e-06 \\
16 Jan 2015, Eclipse 2, \textit{r'} & 0.055 & 0.05 & 0.908 & 0.134 & 0.055 & 0.062 & 0.02 & 0.005 & 57039.1876866 & 3.6e-06 \\
16 Jan 2015, Eclipse 1, \textit{r'} & 0.03 & 0.013 & 0.821 & 0.098 & 0.025 & 0.015 & 0.023 & 0.004 & 57039.153169 & 5.4e-06 \\
16 Jan 2015, Eclipse 3, \textit{g'} & 0.025 & 0.005 & 0.793 & 0.054 & 0.02 & 0.005 & 0.024 & 0.002 & 57039.2221967 & 2.5e-06 \\
16 Jan 2015, Eclipse 4, \textit{g'} & 0.025 & 0.006 & 0.825 & 0.052 & 0.021 & 0.006 & 0.023 & 0.002 & 57039.2567183 & 1.9e-06 \\
16 Jan 2015, Eclipse 1, \textit{g'} & 0.032 & 0.013 & 0.878 & 0.081 & 0.029 & 0.015 & 0.021 & 0.003 & 57039.1531657 & 2.3e-06 \\
16 Jan 2015, Eclipse 5, \textit{u'} & 0.027 & 0.025 & 0.783 & 0.216 & 0.023 & 0.029 & 0.026 & 0.009 & 57039.291242 & 1.56e-05 \\
16 Jan 2015, Eclipse 4, \textit{u'} & 0.026 & 0.025 & 0.807 & 0.221 & 0.024 & 0.032 & 0.025 & 0.009 & 57039.2567093 & 1.21e-05 \\
16 Jan 2015, Eclipse 3, \textit{u'} & 0.022 & 0.011 & 0.64 & 0.143 & 0.015 & 0.01 & 0.032 & 0.007 & 57039.2221917 & 1.87e-05 \\
16 Jan 2015, Eclipse 2, \textit{u'} & 0.029 & 0.036 & 0.736 & 0.208 & 0.026 & 0.047 & 0.028 & 0.008 & 57039.1876878 & 1.48e-05 \\
16 Jan 2015, Eclipse 1, \textit{u'} & 0.033 & 0.041 & 0.692 & 0.205 & 0.028 & 0.05 & 0.03 & 0.009 & 57039.1531702 & 1.58e-05 \\
16 Jan 2015, Eclipse 2, \textit{g'} & 0.023 & 0.005 & 0.818 & 0.048 & 0.019 & 0.005 & 0.023 & 0.002 & 57039.187685 & 1.9e-06 \\
17 Jan 2015, Eclipse 3, \textit{r'} & 0.046 & 0.039 & 0.98 & 0.122 & 0.049 & 0.049 & 0.018 & 0.004 & 57040.2922993 & 3.7e-06 \\
17 Jan 2015, Eclipse 2, \textit{r'} & 0.037 & 0.026 & 0.874 & 0.116 & 0.035 & 0.031 & 0.022 & 0.004 & 57040.257784 & 3.5e-06 \\
17 Jan 2015, Eclipse 1, \textit{r'} & 0.039 & 0.029 & 0.937 & 0.124 & 0.039 & 0.035 & 0.02 & 0.004 & 57040.2232624 & 4e-06 \\
17 Jan 2015, Eclipse 3, \textit{g'} & 0.068 & 0.061 & 1.015 & 0.129 & 0.076 & 0.078 & 0.017 & 0.004 & 57040.2922991 & 2.1e-06 \\
17 Jan 2015, Eclipse 1, \textit{g'} & 0.035 & 0.024 & 0.903 & 0.102 & 0.033 & 0.029 & 0.021 & 0.003 & 57040.2232655 & 2.1e-06 \\
17 Jan 2015, Eclipse 3, \textit{u'} & 0.057 & 0.111 & 0.816 & 0.266 & 0.061 & 0.146 & 0.025 & 0.011 & 57040.2922946 & 1.4e-05 \\
17 Jan 2015, Eclipse 2, \textit{u'} & 0.027 & 0.019 & 0.679 & 0.176 & 0.02 & 0.022 & 0.03 & 0.008 & 57040.2577836 & 1.22e-05 \\
17 Jan 2015, Eclipse 1, \textit{u'} & 0.025 & 0.016 & 0.657 & 0.199 & 0.018 & 0.02 & 0.032 & 0.009 & 57040.2232583 & 1.7e-05 \\
17 Jan 2015, Eclipse 2, \textit{g'} & 0.029 & 0.012 & 0.818 & 0.081 & 0.024 & 0.014 & 0.023 & 0.003 & 57040.257786 & 2e-06 \\
23 May 2015, Eclipse 4, \textit{g'} & 0.039 & 0.018 & 0.942 & 0.09 & 0.039 & 0.022 & 0.019 & 0.003 & 57166.1506635 & 2.2e-06 \\
23 May 2015, Eclipse 6, \textit{u'} & 0.031 & 0.033 & 0.589 & 0.313 & 0.023 & 0.04 & 0.04 & 0.022 & 57166.2197009 & 2.78e-05 \\
23 May 2015, Eclipse 1, \textit{g'} & 0.028 & 0.005 & 0.88 & 0.049 & 0.025 & 0.005 & 0.021 & 0.002 & 57166.0471059 & 2.7e-06 \\
23 May 2015, Eclipse 2, \textit{g'} & 0.026 & 0.004 & 0.848 & 0.045 & 0.023 & 0.005 & 0.022 & 0.002 & 57166.0816193 & 2.2e-06 \\
23 May 2015, Eclipse 3, \textit{g'} & 0.023 & 0.004 & 0.84 & 0.042 & 0.019 & 0.005 & 0.023 & 0.001 & 57166.1161419 & 2e-06 \\
23 May 2015, Eclipse 5, \textit{g'} & 0.028 & 0.01 & 0.921 & 0.07 & 0.027 & 0.012 & 0.02 & 0.002 & 57166.1851826 & 2.2e-06 \\
23 May 2015, Eclipse 5, \textit{u'} & 0.035 & 0.039 & 0.809 & 0.218 & 0.034 & 0.051 & 0.025 & 0.009 & 57166.1851761 & 1.32e-05 \\
23 May 2015, Eclipse 1, \textit{r'} & 0.029 & 0.007 & 0.883 & 0.066 & 0.026 & 0.008 & 0.021 & 0.002 & 57166.0471036 & 5.2e-06 \\
23 May 2015, Eclipse 2, \textit{r'} & 0.027 & 0.007 & 0.856 & 0.069 & 0.024 & 0.008 & 0.022 & 0.002 & 57166.0816159 & 4e-06 \\
23 May 2015, Eclipse 3, \textit{r'} & 0.024 & 0.007 & 0.861 & 0.065 & 0.021 & 0.009 & 0.022 & 0.002 & 57166.1161433 & 4e-06 \\
23 May 2015, Eclipse 4, \textit{r'} & 0.045 & 0.031 & 0.958 & 0.123 & 0.046 & 0.039 & 0.019 & 0.004 & 57166.1506634 & 4.3e-06 \\
23 May 2015, Eclipse 5, \textit{r'} & 0.094 & 0.095 & 1.094 & 0.12 & 0.111 & 0.126 & 0.014 & 0.004 & 57166.1851822 & 3.6e-06 \\
23 May 2015, Eclipse 6, \textit{r'} & 0.026 & 0.009 & 0.886 & 0.094 & 0.024 & 0.011 & 0.021 & 0.003 & 57166.2196978 & 5.8e-06 \\
23 May 2015, Eclipse 6, \textit{g'} & 0.032 & 0.012 & 0.912 & 0.096 & 0.03 & 0.015 & 0.02 & 0.003 & 57166.2197032 & 5.1e-06 \\
23 May 2015, Eclipse 4, \textit{u'} & 0.037 & 0.045 & 0.851 & 0.211 & 0.037 & 0.06 & 0.023 & 0.008 & 57166.1506675 & 1.07e-05 \\
23 May 2015, Eclipse 3, \textit{u'} & 0.022 & 0.01 & 0.78 & 0.166 & 0.018 & 0.012 & 0.025 & 0.007 & 57166.1161449 & 1.26e-05 \\
23 May 2015, Eclipse 2, \textit{u'} & 0.028 & 0.027 & 0.765 & 0.174 & 0.025 & 0.034 & 0.026 & 0.007 & 57166.0816142 & 1.18e-05 \\
23 May 2015, Eclipse 1, \textit{u'} & 0.089 & 0.177 & 0.925 & 0.251 & 0.106 & 0.238 & 0.021 & 0.009 & 57166.0471044 & 1.42e-05 \\

\hline
\end{tabular}
\end{table*}

\begin{table*}
	\centering
	\contcaption{Results from the fits carried out to individual eclipses. $\sigma$ here is the standard deviation of MCMC chain values. We also quote the weighted mean (with its propagated error) and standard deviation of these values for each parameter.}
\begin{tabular}[c]{l c c c c c c c c c c}
\hline
Observation date and filter & $q$ & $\sigma_q$ & $M_1 (M_\odot)$ & $\sigma_{M_1}$ & $M_2 (M_\odot)$ & $\sigma_{M_2}$ & $R_1 / a$ & $\sigma_{R_1/a}$ & $t_0$ (BTDB(MJD)) & $\sigma_{t_0}$ \\
\hline

22 Jun 2015, Eclipse 1, \textit{u'} & 0.024 & 0.018 & 0.668 & 0.156 & 0.017 & 0.02 & 0.03 & 0.007 & 57196.113649 & 1.28e-05 \\
22 Jun 2015, Eclipse 2, \textit{u'} & 0.033 & 0.036 & 0.76 & 0.221 & 0.03 & 0.044 & 0.027 & 0.009 & 57196.1481761 & 1.36e-05 \\
22 Jun 2015, Eclipse 4, \textit{u'} & 0.044 & 0.065 & 0.408 & 0.304 & 0.029 & 0.073 & 0.059 & 0.033 & 57196.2172251 & 3.54e-05 \\
22 Jun 2015, Eclipse 1, \textit{g'} & 0.028 & 0.012 & 0.83 & 0.081 & 0.024 & 0.014 & 0.023 & 0.003 & 57196.1136516 & 2.2e-06 \\
22 Jun 2015, Eclipse 2, \textit{g'} & 0.049 & 0.023 & 0.957 & 0.094 & 0.049 & 0.027 & 0.019 & 0.003 & 57196.1481697 & 2.4e-06 \\
22 Jun 2015, Eclipse 3, \textit{u'} & 0.026 & 0.018 & 0.665 & 0.195 & 0.019 & 0.021 & 0.031 & 0.01 & 57196.1826902 & 1.64e-05 \\
22 Jun 2015, Eclipse 4, \textit{g'} & 0.03 & 0.016 & 0.893 & 0.115 & 0.028 & 0.019 & 0.021 & 0.004 & 57196.2172048 & 6e-06 \\
22 Jun 2015, Eclipse 1, \textit{r'} & 0.034 & 0.019 & 0.863 & 0.109 & 0.031 & 0.023 & 0.022 & 0.004 & 57196.113646 & 4.7e-06 \\
22 Jun 2015, Eclipse 2, \textit{r'} & 0.04 & 0.015 & 0.939 & 0.087 & 0.039 & 0.018 & 0.019 & 0.003 & 57196.1481721 & 4.2e-06 \\
22 Jun 2015, Eclipse 3, \textit{r'} & 0.034 & 0.02 & 0.894 & 0.099 & 0.032 & 0.025 & 0.021 & 0.003 & 57196.1826913 & 4.3e-06 \\
22 Jun 2015, Eclipse 4, \textit{r'} & 0.063 & 0.093 & 1.069 & 0.153 & 0.075 & 0.127 & 0.015 & 0.005 & 57196.2171963 & 6.1e-06 \\
22 Jun 2015, Eclipse 3, \textit{g'} & 0.029 & 0.009 & 0.891 & 0.069 & 0.026 & 0.01 & 0.021 & 0.002 & 57196.1826902 & 2.3e-06 \\
12 Mar 2016, Eclipse 1, \textit{KG5} & 0.031 & 0.006 & 0.939 & 0.303 & 0.029 & 0.011 & 0.021 & 0.011 & 57459.8086767 & 3e-05 \\
13 Mar 2016, Eclipse 1, \textit{KG5} & 0.035 & 0.012 & 0.779 & 0.292 & 0.026 & 0.01 & 0.027 & 0.014 & 57460.8442886 & 8.1e-05 \\
13 Mar 2016, Eclipse 2, \textit{KG5} & 0.035 & 0.017 & 0.798 & 0.365 & 0.027 & 0.014 & 0.028 & 0.017 & 57460.8787594 & 5.01e-05 \\
14 Mar 2016, Eclipse 1, \textit{KG5} & 0.053 & 0.048 & 0.834 & 0.208 & 0.05 & 0.059 & 0.024 & 0.008 & 57461.8798189 & 1.57e-05 \\
14 Mar 2016, Eclipse 2, \textit{KG5} & 0.035 & 0.021 & 0.924 & 0.17 & 0.033 & 0.024 & 0.02 & 0.006 & 57461.9143578 & 1.4e-05 \\
14 Mar 2016, Eclipse 3, \textit{KG5} & 0.041 & 0.025 & 0.915 & 0.283 & 0.035 & 0.022 & 0.021 & 0.011 & 57461.9488529 & 2.93e-05 \\
15 Mar 2016, Eclipse 2, \textit{g'} & 0.041 & 0.026 & 0.763 & 0.195 & 0.033 & 0.025 & 0.026 & 0.008 & 57462.8118723 & 1.64e-05 \\
15 Mar 2016, Eclipse 1, \textit{g'} & 0.035 & 0.025 & 0.862 & 0.22 & 0.032 & 0.028 & 0.023 & 0.009 & 57462.7773383 & 1.58e-05 \\
06 Aug 2016, Eclipse 3, \textit{r'} & 0.029 & 0.022 & 0.787 & 0.155 & 0.025 & 0.026 & 0.025 & 0.006 & 57606.2406934 & 9.7e-06 \\
06 Aug 2016, Eclipse 4, \textit{r'} & 0.032 & 0.023 & 0.898 & 0.118 & 0.031 & 0.028 & 0.021 & 0.004 & 57606.2751919 & 5.1e-06 \\
06 Aug 2016, Eclipse 7, \textit{r'} & 0.03 & 0.022 & 0.874 & 0.118 & 0.028 & 0.028 & 0.022 & 0.004 & 57606.3787416 & 4.9e-06 \\
06 Aug 2016, Eclipse 6, \textit{r'} & 0.047 & 0.046 & 0.92 & 0.163 & 0.049 & 0.059 & 0.02 & 0.006 & 57606.3442234 & 7.8e-06 \\
06 Aug 2016, Eclipse 2, \textit{r'} & 0.024 & 0.007 & 0.83 & 0.069 & 0.021 & 0.009 & 0.023 & 0.002 & 57606.2061493 & 3.7e-06 \\
06 Aug 2016, Eclipse 5, \textit{r'} & 0.022 & 0.006 & 0.756 & 0.068 & 0.017 & 0.006 & 0.026 & 0.003 & 57606.3097068 & 5e-06 \\
06 Aug 2016, Eclipse 1, \textit{r'} & 0.024 & 0.009 & 0.81 & 0.098 & 0.02 & 0.011 & 0.024 & 0.004 & 57606.1716347 & 5.7e-06 \\
06 Aug 2016, Eclipse 7, \textit{g'} & 0.027 & 0.015 & 0.832 & 0.094 & 0.024 & 0.018 & 0.023 & 0.003 & 57606.3787506 & 3.9e-06 \\
06 Aug 2016, Eclipse 6, \textit{g'} & 0.039 & 0.055 & 0.864 & 0.148 & 0.04 & 0.072 & 0.022 & 0.005 & 57606.344228 & 6.4e-06 \\
06 Aug 2016, Eclipse 5, \textit{g'} & 0.03 & 0.016 & 0.888 & 0.103 & 0.028 & 0.02 & 0.021 & 0.003 & 57606.3097093 & 4.1e-06 \\
06 Aug 2016, Eclipse 4, \textit{g'} & 0.025 & 0.011 & 0.835 & 0.085 & 0.021 & 0.013 & 0.023 & 0.003 & 57606.2751921 & 4.2e-06 \\
06 Aug 2016, Eclipse 3, \textit{g'} & 0.034 & 0.037 & 0.896 & 0.155 & 0.035 & 0.048 & 0.021 & 0.005 & 57606.2406716 & 7.6e-06 \\
06 Aug 2016, Eclipse 2, \textit{g'} & 0.026 & 0.007 & 0.823 & 0.064 & 0.021 & 0.008 & 0.023 & 0.002 & 57606.2061493 & 3e-06 \\
06 Aug 2016, Eclipse 1, \textit{g'} & 0.023 & 0.006 & 0.833 & 0.063 & 0.02 & 0.007 & 0.023 & 0.002 & 57606.171626 & 3.6e-06 \\
07 Aug 2016, Eclipse 6, \textit{r'} & 0.03 & 0.019 & 0.858 & 0.119 & 0.027 & 0.023 & 0.022 & 0.004 & 57607.3452922 & 6e-06 \\
07 Aug 2016, Eclipse 5, \textit{r'} & 0.032 & 0.024 & 0.761 & 0.122 & 0.027 & 0.029 & 0.026 & 0.005 & 57607.3107731 & 6.6e-06 \\
07 Aug 2016, Eclipse 4, \textit{r'} & 0.038 & 0.033 & 0.845 & 0.139 & 0.036 & 0.041 & 0.023 & 0.005 & 57607.2762623 & 5.2e-06 \\
07 Aug 2016, Eclipse 3, \textit{r'} & 0.027 & 0.012 & 0.863 & 0.103 & 0.025 & 0.014 & 0.022 & 0.004 & 57607.2417342 & 6.1e-06 \\
07 Aug 2016, Eclipse 1, \textit{r'} & 0.028 & 0.015 & 0.757 & 0.103 & 0.023 & 0.016 & 0.026 & 0.004 & 57607.1726877 & 6.5e-06 \\
07 Aug 2016, Eclipse 6, \textit{g'} & 0.032 & 0.019 & 0.859 & 0.105 & 0.029 & 0.022 & 0.022 & 0.004 & 57607.3452998 & 4.6e-06 \\
07 Aug 2016, Eclipse 2, \textit{r'} & 0.026 & 0.013 & 0.806 & 0.086 & 0.022 & 0.014 & 0.024 & 0.003 & 57607.2072221 & 4.3e-06 \\
07 Aug 2016, Eclipse 4, \textit{g'} & 0.033 & 0.03 & 0.873 & 0.118 & 0.032 & 0.037 & 0.022 & 0.004 & 57607.2762599 & 4.3e-06 \\
07 Aug 2016, Eclipse 3, \textit{g'} & 0.028 & 0.013 & 0.878 & 0.091 & 0.025 & 0.016 & 0.021 & 0.003 & 57607.2417351 & 4.2e-06 \\
07 Aug 2016, Eclipse 2, \textit{g'} & 0.023 & 0.005 & 0.818 & 0.056 & 0.019 & 0.005 & 0.023 & 0.002 & 57607.2072187 & 2.7e-06 \\
07 Aug 2016, Eclipse 1, \textit{g'} & 0.027 & 0.013 & 0.803 & 0.095 & 0.023 & 0.015 & 0.024 & 0.003 & 57607.1726871 & 4e-06 \\
07 Aug 2016, Eclipse 5, \textit{g'} & 0.028 & 0.013 & 0.867 & 0.088 & 0.025 & 0.015 & 0.022 & 0.003 & 57607.3107734 & 4e-06 \\
08 Aug 2016, Eclipse 4, \textit{r'} & 0.028 & 0.016 & 0.78 & 0.1 & 0.023 & 0.014 & 0.025 & 0.004 & 57608.2773267 & 5.2e-06 \\
08 Aug 2016, Eclipse 1, \textit{g'} & 0.034 & 0.021 & 0.855 & 0.107 & 0.031 & 0.025 & 0.022 & 0.004 & 57608.1737669 & 4e-06 \\
08 Aug 2016, Eclipse 2, \textit{g'} & 0.027 & 0.015 & 0.837 & 0.097 & 0.024 & 0.018 & 0.023 & 0.003 & 57608.2082868 & 3e-06 \\
08 Aug 2016, Eclipse 3, \textit{g'} & 0.027 & 0.014 & 0.802 & 0.09 & 0.023 & 0.016 & 0.024 & 0.003 & 57608.2427987 & 3e-06 \\
08 Aug 2016, Eclipse 4, \textit{g'} & 0.026 & 0.013 & 0.828 & 0.087 & 0.022 & 0.012 & 0.023 & 0.003 & 57608.2773237 & 3.1e-06 \\
08 Aug 2016, Eclipse 6, \textit{g'} & 0.034 & 0.018 & 0.846 & 0.108 & 0.03 & 0.021 & 0.023 & 0.004 & 57608.3463644 & 3.4e-06 \\
08 Aug 2016, Eclipse 1, \textit{r'} & 0.031 & 0.019 & 0.839 & 0.115 & 0.028 & 0.023 & 0.023 & 0.004 & 57608.1737637 & 5.6e-06 \\
08 Aug 2016, Eclipse 2, \textit{r'} & 0.033 & 0.019 & 0.878 & 0.11 & 0.03 & 0.023 & 0.021 & 0.004 & 57608.2082847 & 4.8e-06 \\
08 Aug 2016, Eclipse 3, \textit{r'} & 0.03 & 0.019 & 0.895 & 0.104 & 0.028 & 0.023 & 0.021 & 0.003 & 57608.2428044 & 3.6e-06 \\
08 Aug 2016, Eclipse 6, \textit{r'} & 0.03 & 0.014 & 0.852 & 0.1 & 0.027 & 0.016 & 0.022 & 0.003 & 57608.3463683 & 5.4e-06 \\
21 Feb 2017, Eclipse 1, \textit{KG5} & 0.042 & 0.041 & 0.771 & 0.223 & 0.039 & 0.052 & 0.026 & 0.009 & 57805.8673572 & 1.45e-05 \\

\hline
\textbf{Weighted Mean} & 0.030 & 0.003 & 0.808 & 0.074 & 0.025 & 0.002 & 0.022 & 0.002 & - & - \\
\textbf{Standard Dev.} & 0.014 & - & 0.092 & - & 0.017 & - & 0.005 & - & - & - \\
\hline
\hline
Phase-fold, \fr & 0.0296 & 0.0008 & 0.87 & 0.06 & 0.026 & 0.002 & 0.021 & 0.002 & - & - \\
Phase-fold, \fg & 0.0312 & 0.0007 & 0.90 & 0.06 & 0.028 & 0.002 & 0.021 & 0.002 & - & - \\
\textbf{Weighted Mean} & 0.0305 & 0.0005 & 0.90 & 0.04 & 0.0273 & 0.0014 & 0.0207 & 0.0014 & - & - \\
\hline
\end{tabular}
\end{table*}


\bsp	
\label{lastpage}
\end{document}